\documentclass[a4paper]{article}

\usepackage{graphicx}
\usepackage{amsmath,amssymb}
\usepackage{url}
\usepackage{colonequals} % defines nice :- as \colonminus  and := as \colonequals
\usepackage{natbib}

\newcommand{\coloneq}{\mbox{ :- }}
\newcommand{\true}{\ensuremath{\mathrm{true}}}
\newcommand{\false}{\ensuremath{\mathrm{false}}}

\newcommand{\clpbn}
  {CLP($\mathcal{BN}$)}
\newcommand{\val}{\ensuremath{\simeq\!\!}}

\begin{document}

\title{Probabilistic Programming Concepts}

\author{Luc De Raedt ~~~~~ Angelika Kimmig \\ \small{Department of Computer Science, KU Leuven}\\
       \small{Celestijnenlaan 200a -- box 2402, 3001 Heverlee, Belgium}\\
              \small{\texttt{\{firstname.lastname\}@cs.kuleuven.be}}    
}

\date{}

\maketitle

\begin{abstract}
A multitude of different probabilistic programming languages exists
today, all extending a traditional programming language with
primitives to support modeling of complex, structured probability distributions. 
Each of these languages employs its own probabilistic primitives, and comes with a particular syntax, semantics 
and inference procedure.  This makes it hard to understand the underlying programming concepts
and appreciate the differences between the different languages. 

To obtain a better understanding of probabilistic programming, we 
identify a number of core programming concepts underlying the primitives used by various probabilistic languages,
discuss the execution mechanisms that they require and use these to position state-of-the-art probabilistic 
languages and their implementation.  

While doing so, we focus on probabilistic extensions of {\em logic} programming languages such as Prolog, 
which have been developed since more than 20 years.
\end{abstract}

\section{Introduction}
\label{sec:intro}
The vast interest in statistical relational learning \citep{Getoor07:book}, probabilistic (inductive) logic programming \citep{DeRaedt07-APRIL} and probabilistic programming languages \citep{GoodmannNIPSWS} has resulted in a wide variety of different formalisms, models and languages. 
The multitude of probabilistic languages that exists today provides evidence for the richness and maturity of the field, but on the other hand, makes it hard to get an appreciation and understanding of the relationships and differences between the different languages. 
Furthermore, most arguments in the literature about the relationship amongst these languages are about the 
expressiveness of these languages, that is, they state (often in an informal way) that one language is more expressive than another one (implying that the former could be used to emulate the latter).   By now, it is commonly accepted that 
the more interesting question is concerned with the underlying concepts that these languages employ
and their effect on the inference mechanisms, as their expressive power is often very similar.   However, 
a multitude of different probabilistic primitives exists, which makes it hard to 
appreciate their relationships.\footnote{Throughout the paper we use the term \emph{primitive} to denote a particular syntactic and semantic
construct that is available in a particular probabilistic programming
language, and the term \emph{concept} to denote the 
underlying notion. Different primitives may hence realize the same concept. }

To alleviate these difficulties and obtain a better understanding of the field
we identify a number of core probabilistic programming concepts and
relate them to one another. We cover the basic
concepts representing different types of random variables, but also
general modeling concepts such as negation or time and dynamics, and
programming constructs such as meta-calls and ways to handle sets. 
While doing so, we focus on probabilistic extensions of logic programming languages
because this is (arguably) the first and best studied probabilistic programming paradigm. 
It has been studied for over 20 years starting with the seminal work of David \cite{Poole92} and Taisuke \cite{Sato:95},
and now includes languages such as  \clpbn~\citep{clpbn},
BLPs~\citep{Kersting08}, ICL~\citep{Poole08},
PRISM~\citep{SatoKameya:01}, ProbLog~\citep{DeRaedt07-IJCAIa},
LPADs~\citep{Vennekens04}, 
CP-logic \citep{vennekens:tplp09},  SLPs \citep{Muggleton:96},
PROPPR~\citep{wang:cikm13}, P-log \citep{baral:tplp09} and Dyna \citep{Eisner05}. 
Another reason for focussing on probabilistic extensions of logic programming languages
is that the concepts are all embedded within the same host language, so we can focus on semantics rather than syntax. 
At the same time, we also relate the concepts to alternative
probabilistic programming languages such as Church \citep{Goodman08},
IBAL \citep{Pfeffer01}, Figaro \citep{Pfeffer09} and BLOG \citep{Milch05}
and to some extent also to statistical relational learning models such
as RBNs \citep{Jaeger08}, Markov logic \citep{Richardson:06},
 and PRMs \citep{getoor}.  Most statistical relational learning approaches
employ a knowledge-based model construction approach, in which the
logic is used as a template
for constructing a graphical model. Typical probabilistic programming
languages, on the other hand, employ a variant of Sato's distribution semantics \citep{Sato:95}, in which 
random variables directly correspond to ground facts and a traditional program 
specifies how to deduce further knowledge from these facts. This
difference explains
why we introduce the concepts in the context of the distribution semantics,
and discuss approaches to knowledge-based model construction separately.

Inference is a key challenge in probabilistic programming and
statistical relational learning.  
Furthermore, the choice of inference approach often influences which
probabilistic primitives can be supported. 
Enormous progress has been made in the past few years w.r.t. probabilistic inference and numerous inference procedures have been contributed. 
Therefore, we also identify some core classes of inference mechanisms for probabilistic programming and discuss which 
ones to use for which probabilistic concept. Inference in probabilistic languages also
is an important building block of approaches that learn the structure
and/or parameters of such models from data. Given the variety of
approaches that exist today, a discussion of learning is beyond the
scope of this paper. 

To summarize, the key contributions of this paper are (1) 
the identification of a number of core concepts that are used by various probabilistic languages, (2) 
a discussion of the execution mechanisms that they require, 
and (3) a positioning of state-of-the-art probabilistic languages and implementations 
w.r.t. these concepts.  
Although many of the concepts we discuss are well-described in the literature, some even in survey papers \citep{deraedt:sigkdd03,Poole08}, 
we believe a new and up-to-date survey is warranted due to the rapid developments
of the field which rapidly renders existing surveys incomplete and even outdated. 
To the best of our knowledge, this 
is also the first time that such a wide variety of probabilistic programming 
concepts and languages, also in connection to inference, is discussed in a single paper.

We expect the reader to be familiar with basic language concepts and
terms of Prolog~\citep{lloyd:book89,Fla94:book}; a quick summary can be found
in Appendix~\ref{app:lp}.

This paper is organized as follows. 
We first discuss the distribution semantics (Section~\ref{sec:ds-bg})
and classify corresponding inference approaches according to their
logical and probabilistic components 
(Section~\ref{sec:inference}). Section~\ref{sec:concepts} identifies
the probabilistic programming concepts. In Section~\ref{sec:kbmc-bg}, we
discuss the relation with statistical relational modeling approaches rooted in graphical
models. Section~\ref{sec:ppci} relates the different inference
approaches to the probabilistic programming concepts.

\section{Distribution Semantics}\label{sec:ds-bg}
Sato's distribution semantics~\citep{Sato:95} is a well-known
semantics for probabilistic logics that has been used many times in
the literature,
cf.~\citep{Dantsin,Poole:93,Fuhr00,Poole00,SatoKameya:01,DalviS04,DeRaedt07-IJCAIa}.
Prominent examples of Prolog-based languages using this semantics
include 
 ICL~\citep{Poole08},
PRISM~\citep{SatoKameya:01} and
ProbLog~\citep{DeRaedt07-IJCAIa,Kimmig11}, even though
there exist subtle differences between these languages as we will
illustrate later.
Sato has defined the distribution
semantics  for a countably infinite set of random variables and a
general class of distributions. We focus on the finite case here,
discussing the two most popular instances of the semantics, based on
a set of independent random variables and independent probabilistic
choices, respectively,  and refer
to \citep{Sato:95} for details on the general case.

\subsection{Probabilistic Facts}
The arguably most basic instance of the distribution semantics uses a
finite set of Boolean random variables that are all pairwise
independent. Throughout the paper, we use the following running example inspired
by the well-known alarm Bayesian network: 
\begin{align}
0.1::\mathtt{burglary.} ~~~~&0.7::\mathtt{hears\_alarm(mary).} \nonumber\\
0.2::\mathtt{earthquake.} ~~~~&0.4::\mathtt{hears\_alarm(john).}\nonumber\\
\mathtt{alarm}&\colonminus\mathtt{earthquake.}\label{ex:alarm}\\
\mathtt{alarm}&\colonminus\mathtt{burglary.}\nonumber\\
\mathtt{calls(X)}&\colonminus\mathtt{alarm, hears\_alarm(X).}\nonumber\\
\mathtt{call}&\colonminus\mathtt{calls(X).}\nonumber
\end{align}\label{page:alarm}
The program consists of a set $R$ of definite clauses or \emph{rules},
and a set $F$ of ground facts $f$, each of them
labeled with a probability $p$, written as $p::f$. We call such
labeled facts \emph{probabilistic facts}. Each probabilistic fact
corresponds to a Boolean random
variable that is \emph{true} with probability $p$ and \emph{false} with
probability $1-p$. We use $b$, $e$, $hm$ and $hj$ to denote the random
variables corresponding to \verb|burglary|, \verb|earthquake|,
\verb|hears_alarm(mary)| and \verb|hears_alarm(john)|, respectively. 
Assuming that all these random variables are
independent, we obtain the following probability distribution $P_F$
over truth value assignments to these random variables and their
corresponding  sets
of ground facts $F'\subseteq F$:
\begin{equation}
\label{eq:p_f}
P_F(F') = \prod_{f_i\in F'}p_i\cdot\prod_{f_i\in F\setminus F'}(1-p_i)
\end{equation}
For instance, the truth value assignment $\mathtt{burglary}=\true$,
$\mathtt{earthquake}=\false$, $\mathtt{hears\_alarm(mary)}=\true$, $\mathtt{hears\_alarm(john)}=\false$, which we will abbreviate as $b\wedge \neg e\wedge
hm\wedge\neg hj$, corresponds to the set of facts $\{$\verb|burglary|,
\verb|hears_alarm(mary)|$\}$, and has probability $0.1\cdot
(1-0.2)\cdot 0.7 \cdot (1-0.6) = 0.0336$. The corresponding logic
program obtained by adding the set of rules $R$ to the set of facts,
also called a \emph{possible world}, is 
\begin{align}
\mathtt{burglary.}&\nonumber\\
\mathtt{hears\_alarm(mary).}& \nonumber\\
\mathtt{alarm}&\colonminus\mathtt{earthquake.}\label{ex:possworld}\\
\mathtt{alarm}&\colonminus\mathtt{burglary.}\nonumber\\
\mathtt{calls(X)}&\colonminus\mathtt{alarm, hears\_alarm(X).}\nonumber\\
\mathtt{call}&\colonminus\mathtt{calls(X).}\nonumber
\end{align}
As each logic program obtained by fixing the truth values of all
probabilistic facts has a unique least Herbrand model, $P_F$ can be
used to define the \emph{success probability} of a query $q$, that is,
the probability that $q$ is true in a randomly chosen such program,
as the sum over all programs that entail $q$:
\begin{align}
\label{eq:p_suc}
P_s(q) &\colonequals \sum\limits_{\substack{F'\subseteq F \\
    \exists\theta F'\cup R\models q\theta}} P_F(F')\\
 &= \sum\limits_{\substack{F'\subseteq F \\
    \exists\theta F'\cup R\models q\theta}} \prod_{f_i\in F'}p_i\cdot\prod_{f_i\in F\setminus F'}(1-p_i)\;.\label{eq:p_suc_facts}
\end{align}
Naively, the success probability can thus be computed by enumerating
all sets $F'\subseteq F$, for each of them checking whether
the corresponding possible world entails the query, and summing the
probabilities of those that do. As fixing the set of facts yields an
ordinary logic program, the entailment check can use any reasoning
technique for such programs.
 
For instance, \emph{forward reasoning}, also
known as applying the $T_P$ operator, starts from the set of
facts and repeatedly uses rules to derive additional facts until no more facts
can be derived. In our example possible world \eqref{ex:possworld}, we thus start 
from $\{$\verb|burglary|, \verb|hears_alarm(mary)|$\}$, and first add \verb|alarm| due to the
second rule based on \verb|burglary|. This in turn makes it possible
to add \verb|calls(mary)| using the third rule and substitution
\verb|X|=\verb|mary|, and finally, \verb|call| is added using the last rule,
resulting in the least Herbrand model $\{$\verb|burglary|,
\verb|hears_alarm(mary)|, \verb|alarm|, \verb|calls(mary)|,
\verb|call|$\}$. This possible world thus contributes to the success
probabilities of \verb|alarm|, \verb|calls(mary)| and \verb|call|, but
not to the one of \verb|calls(john)|. 

An alternative to forward reasoning is \emph{backward reasoning}, also
known as SLD-resolution or proving, which we again illustrate for our
example possible world \eqref{ex:possworld}. \label{bw-example} It 
starts from a given query, e.g., \verb|call|, and uses the rules in
the opposite direction: in order to prove a fact appearing in the head
of a clause, we have to prove all literals in the clause's body. For
instance, based on the last rule, to prove \verb|call|, we need to
prove \verb|calls(X)| for some instantiation of \verb|X|. Using the
third rule, this means proving \verb|alarm, hears_alarm(X)|. To prove
\verb|alarm|, we could use the first rule and prove \verb|earthquake|,
but this fails for our choice of facts, as there is no rule (or fact) for the
latter. We thus \emph{backtrack} to the second rule for \verb|alarm|,
which requires proving \verb|burglary|, which is proven by the
corresponding fact. Finally, we prove \verb|hears_alarm(X)| using the
fact \verb|hears_alarm(mary)|, substituting \verb|mary| for \verb|X|,
which completes the proof for \verb|call|. 

Going over all possible worlds
in this way, we obtain the success probability of
\verb|calls(mary)|, $P_s(\mathtt{calls(mary)}) = 0.196$, as the sum of the probabilities of six possible
worlds (listed in Table~\ref{tab:marycalls}).
\begin{table}[t]
\centering
\begin{tabular}{l|c|r}
 world &  \verb|calls(john)|  &   probability\\\hline
 $b\wedge\neg e \wedge hm\wedge\neg hj$ &  \false   & $0.1\cdot (1-0.2)\cdot 0.7\cdot (1-0.4) = 0.0336$ \\
 $b\wedge\neg e \wedge hm \wedge hj$ & \true  &  $0.1\cdot (1-0.2)\cdot 0.7\cdot 0.4 = 0.0224$ \\
 $b\wedge e\wedge hm\wedge\neg hj$ &  \false   & $0.1\cdot 0.2\cdot 0.7\cdot (1-0.4) = 0.0084$ \\
$b\wedge e\wedge hm\wedge hj$  &  \true  & $0.1\cdot 0.2\cdot 0.7\cdot 0.4 = 0.0056$ \\
 $\neg b\wedge e\wedge hm\wedge\neg hj$ & \false   &  $(1-0.1)\cdot 0.2\cdot 0.7\cdot (1-0.4) = 0.0756$\\
 $\neg b\wedge e\wedge hm\wedge hj$ &  \true   & $(1-0.1)\cdot 0.2\cdot 0.7\cdot 0.4 = 0.0504$
\end{tabular}
\caption{The possible worlds of program \eqref{ex:alarm} where
  \texttt{calls(mary)} is true. }
\label{tab:marycalls}
\end{table}

Clearly, enumerating all possible worlds is infeasible for larger
programs; we will discuss alternative inference techniques from the
literature in Section~\ref{sec:inference}.

For ease of modeling (and to allow for countably infinite sets of
probabilistic facts), probabilistic languages such as ICL and ProbLog   use \emph{non-ground probabilistic facts} to define sets of random variables. All ground instances of such a fact are mutually independent and share the same probability value. 
As an example, consider a simple coin game which can be won either by throwing two times heads or by cheating. This game can be modeled by the program below. The probability to win the game is then defined by the success probability $P_s(\mathtt{win})$.
\begin{equation*}
\begin{array}{lll}
0.5 :: \mathtt{heads(X).} & ~~~~ & 0.2 :: \mathtt{cheat\_successfully.} \\
\mathtt{win \colonminus cheat\_successfully.}\\
\mathtt{win \colonminus heads(1), heads(2).}
\end{array}
\end{equation*}
Legal groundings of such facts can also be restricted by providing a
domain, as in the following variant of our alarm
example where all persons have the same probability of independently hearing
the alarm:
\begin{align*}
0.1::\mathtt{burglary.} ~~~~&0.2::\mathtt{earthquake} \\
0.7::\mathtt{hears\_alarm(X)}&\colonminus\mathtt{person(X).}\\
\mathtt{person(mary).~~~} &\mathtt{ person(john).  ~~~person(bob).  ~~~person(ann).}\\
\mathtt{alarm}&\colonminus\mathtt{earthquake.}\\
\mathtt{alarm}&\colonminus\mathtt{burglary.}\\
\mathtt{calls(X)}&\colonminus\mathtt{alarm, hears\_alarm(X).}\\
\mathtt{call}&\colonminus\mathtt{calls(X).}
\end{align*}
If such domains are defined purely logically, without using
probabilistic facts, the basic distribution is still
well defined.

It is often assumed that probabilistic facts do not unify with other
probabilistic facts or heads of rules. 

\subsection{Probabilistic Choices}
\label{sec:ads}
As already noted by \cite{Sato:95}, probabilistic facts (or binary
switches) are expressive enough to represent a wide range of models,
including Bayesian networks, Markov chains and hidden Markov
models. However, for ease of modeling, it is often more convenient to
use multi-valued random variables instead of binary ones. The concept
commonly used to realize such variables in the distribution semantics
is a  probabilistic choice, that is, a finite 
set of ground atoms exactly one of which is true in any possible
world. 
Examples of such choices are the \emph{probabilistic alternatives} of the
Independent Choice Logic (ICL)~\citep{Poole00} and probabilistic Horn
abduction (PHA)~\citep{Poole:93}, the \emph{multi-ary random switches}
of PRISM~\citep{SatoKameya:01}, the \emph{probabilistic clauses} of
stochastic logic programs (SLPs)~\citep{Muggleton:96}, and the
\emph{annotated disjunctions} of logic programs with annotated
disjunctions (LPADs)~\citep{Vennekens04}, or the \emph{CP-events}
of CP-logic~\citep{Vennekens07}. These are all closely related, e.g.,
the probabilistic clauses of SLPs map onto the switches of PRISM
\citep{cussens:pr05}, and the probabilistic alternatives of ICL onto
annotated disjunctions (and vice versa) \citep{Vennekens04}. We
therefore  restrict the following discussion
to annotated disjunctions~\citep{Vennekens04}, using the notation
introduced below.

An \emph{annotated disjunction} (AD) is an expression of the form 
\begin{equation*}
\mathtt{p_1::h_1 ;\ \ldots\ ; p_N::h_N \colonminus b_1,\ \ldots\ ,b_M.}
\end{equation*}
where $\mathtt{b_1, \ldots , b_M}$ is a possibly empty
conjunction of literals, the $p_i$ are probabilities and $\sum_{i=1}^N
p_i\le1$. Considered in isolation, an annotated disjunction states that if the body $\mathtt{b_1,
\ldots , b_M}$ is true at most one of the $\mathtt{h_i}$ is true as
well, where
the choice is governed by the probabilities (see below for
interactions between multiple ADs with unifying atoms in the head). If the $p_i$ in an annotated disjunction
do not sum to 1, there is also the case that nothing is chosen. The
probability of this event is $1-\sum_{i=1}^n p_i$. A probabilistic
fact is thus a special case of an AD with a single head atom and empty
body. 

For instance, consider the following program:
\begin{align*}
& 0.4::\mathtt{draw}.\\
& \frac{1}{3}::\mathtt{color(green)} ; \frac{1}{3}::\mathtt{color(red)}
 ; \frac{1}{3}::\mathtt{color(blue)} \colonminus \mathtt{draw}.
\end{align*}
The probabilistic fact states that we draw a ball from an urn with
probability $0.4$, and the annotated disjunction states that if we
draw a ball, the color is picked uniformly among \verb|green|,
\verb|red| and \verb|blue|. As for probabilistic facts,
a non-ground AD denotes the set of all its groundings, and for each
such grounding, choosing one of its head atoms to be true is seen as
an independent random event. That is, the annotated disjunction
\begin{equation*}
 \frac{1}{3}::\mathtt{color(B,green)} ;
 \frac{1}{3}::\mathtt{color(B,red)} ;
 \frac{1}{3}::\mathtt{color(B,blue) \colonminus ball(B).}
\end{equation*}
defines an independent probabilistic choice of color for each ball
\verb|B|. 

As noted already by \citet{Vennekens04}, the probabilistic choice over
head atoms in an annotated disjunction can equivalently be expressed
using a set of logical clauses, one for each head, and a probabilistic
choice over facts added to the bodies of these clauses, e.g.
\begin{align*}
\mathtt{color(B,green) \colonminus ball(B), choice(B,green).}\\
\mathtt{color(B,red) \colonminus ball(B), choice(B,red).}\\
\mathtt{color(B,blue) \colonminus ball(B), choice(B,blue).}\\
 \frac{1}{3}::\mathtt{choice(B,green)} ;
 \frac{1}{3}::\mathtt{choice(B,red)} ;
 \frac{1}{3}::\mathtt{choice(B,blue)}. 
\end{align*}
This example illustrates that annotated disjunctions define  a distribution  $P_F$
over basic facts as required in the 
distribution semantics, but can simplify modeling by directly expressing
probabilistic consequences. 

As mentioned above, a probabilistic fact directly corresponds to an
annotated disjunction with a single atom in the head and an empty
body. Conversely, each annotated disjunction can -- for the purpose of
calculating success probabilities -- be equivalently
represented using a set of probabilistic facts and deterministic
clauses, which together simulate a sequential choice mechanism; we
refer to Appendix~\ref{app:adpf} for details.

\paragraph{Independent Causes}\label{sec:noisy-or-in-lp} 
Some languages, e.g.~ICL~\citep{Poole08}, assume that head atoms in
the  same or different annotated disjunctions cannot unify with one
another, while others, e.g., LPADs~\citep{Vennekens04}, do not make
this restriction, but instead view each annotated disjunction as an 
independent cause for the conclusions to hold. In that case, the
structure of the program defines the combined effect of these causes,
similarly to how the two clauses for \verb|alarm| in our earlier
example \eqref{ex:alarm} combine the two causes \verb|burglary| and \verb|earthquake|.  
We illustrate this on the Russian roulette example by \cite{vennekens:tplp09},
which involves two guns.
\begin{equation*}
\begin{array}{ll}
\frac{1}{6}::\mathtt{death}  \mathtt{\colonminus} & \mathtt{pull\_trigger(left\_gun).}\\
\frac{1}{6}::\mathtt{death}  \mathtt{\colonminus} & \mathtt{pull\_trigger(right\_gun).}\\
\end{array}
\end{equation*}
Each gun is an independent cause for death. Pulling both triggers will
result in death being true with a probability of $1 -(1-\frac{1}{6})^2$,
which exactly corresponds to the probability of \verb|death| being
proven via the first or via the second annotated disjunction (or
both). 
Assuming independent causes closely corresponds to the noisy-or combining rule 
that is often employed in the Bayesian network literature, cf. Section
\ref{sec:kbmc-bg}.

\subsection{Inference Tasks}\label{sec:tasks}
In probabilistic programming and statistical relational learning, the following inference tasks have been considered:
\begin{itemize}
\item In the $SUCC(q)$ task, a ground query $q$ is given, and the task is to compute
$$SUCC(q) = P_s(q),$$ the success probability of the query as specified in Equation~\eqref{eq:p_suc}.\footnote{Non-ground queries have 
    been considered as well, in which case the success probability
    corresponds to the probability that $q\theta$ is true for some
    grounding substitution $\theta$.}
\item In the $MARG(Q \mid e)$ task, a set $Q$ of ground atoms of interest, the query atoms,
and a ground query $e$, the evidence, are given. 
The task is to compute the marginal probability distribution of each
atom $q\in Q$ given the evidence, 
$$P_s(q|e)  =  {{P_s(q \wedge e)}\over{P_s(e)}}.$$
The $SUCC(q)$ task corresponds to the special case of 
the $MARG(Q\mid e)$ task with $Q=\{q\}$ and $e=\true$ (and
thus $P_s(e)=1$).
\item
The $MAP(Q \mid e)$ task is to find the most likely truth-assignment $q$ to the 
atoms in $Q$ given the evidence $e$, that is, to compute
$$MAP(Q \mid e) = \arg\max_q P_s(Q=q |e)$$ 
\item
The $MPE(U \mid e)$ task is to find the most likely world where the given
evidence query $e$ holds. Let $U$ be the set of all atoms in the
Herbrand base that do not occur in $e$. Then, the task is to compute 
the most likely truth-assignment $u$ to the atoms in $U$,
$$MPE(e) = MAP(U \mid e).$$
\item
In the $VIT(q)$ task, a query $q$ is given, and the task is to find
a Viterbi proof of~$q$. Let $E(q)$ be the set of all explanations or
proofs of
$q$, that is, of all sets $F'$ of ground probabilistic atoms for which
$q$ is true in the corresponding possible world. Then, the task is to
compute
$$VIT(q) = \arg\max_{X\in E(q)} P_s(\bigwedge_{f \in  X} f).$$
\end{itemize}
To illustrate, consider our initial alarm example \eqref{ex:alarm} with 
$e=\mathtt{calls(mary)}$ and 
$Q=\{\mathtt{burglary}, \mathtt{calls(john)}\}$. The worlds  where the
evidence holds are listed in Table~\ref{tab:marycalls}, together with
their probabilities. 
The answer to the
MARG 
task 
is $P_s(\mathtt{burglary}|\mathtt{calls(mary)}) = 0.07/0.196 = 0.357$ and
$P_s(\mathtt{calls(john)}|\mathtt{calls(mary)}) = 0.0784/0.196=0.4$. The answer to the MAP
task is \verb|burglary|=\false, \verb|calls(john)|=\false, as its
probability $0.0756/0.196$ is higher than $0.028/0.196$ (for \true, \true), $0.042/0.196$
(for \true, \false) and $0.0504/0.196$ (for \false, \true). The world returned by MPE is the one corresponding
to the set of facts \verb|{earthquake, hears_alarm(mary)}|. Finally,
the Viterbi proof of query \verb|calls(mary)| is $e\wedge hm$, as $0.2\cdot
0.7 > 0.1\cdot 0.7$ (for $b\wedge hm$).

\section{Inference}
\label{sec:inference} 
We now provide an overview of existing inference approaches in
probabilistic (logic) programming. As most existing work adresses the
SUCC task of computing success probabilities, 
cf.~Equation~\eqref{eq:p_suc}, we focus on this task here, and mention
other tasks in passing where appropriate. For simplicity, we assume
probabilistic facts as basic building blocks. 
Computing marginals under the distribution semantics has to take
into account both \emph{probabilistic} and \emph{logical} aspects. We
therefore distinguish between \emph{exact}
inference and approximation using either \emph{bounds} or
\emph{sampling} on the probabilistic side, and between methods based
on \emph{forward} and \emph{backward reasoning} and \emph{grounding to
CNF} on the logical side. 
Systems implementing (some of) these approaches include the ICL system
AILog2\footnote{\url{http://artint.info/code/ailog/ailog2.html}}, the 
PRISM system\footnote{\url{http://sato-www.cs.titech.ac.jp/prism/}\label{foot:prism}}, the
ProbLog implementations 
ProbLog1\footnote{included in YAP Prolog,
  \url{http://www.dcc.fc.up.pt/~vsc/Yap/}\label{foot:p1}} and
ProbLog2\footnote{\url{http://dtai.cs.kuleuven.be/problog/}\label{foot:p2}}, and the LPAD implementations cplint\footnote{included in YAP Prolog,
  \url{http://www.dcc.fc.up.pt/~vsc/Yap/}} and PITA\footnote{included
  in XSB Prolog, \url{http://xsb.sourceforge.net/}}. General
statements about systems in the following refer to these six systems.

\subsection{Exact Inference}\label{sec:exact}
As most methods for exact inference can be viewed as operating
(implicitly or explicitly) on a propositional logic representation of
all possible worlds that entail the query $q$ of interest, we first note
that this set of possible worlds is given by the following 
formula in disjunctive normal form (DNF)
\begin{equation}
\label{eq:model_dnf}
DNF(q) = \bigvee_{\substack{F'\subseteq F \\
    \exists\theta F'\cup R\models q\theta}} \left(\bigwedge_{f_i\in
  F'}f_i\wedge\bigwedge_{f_i\in F\setminus F'}\neg f_i\right) 
\end{equation}  
and that the structure of this formula exactly mirrors that of
Equation~\eqref{eq:p_suc_facts} defining  the
success probability in the case of probabilistic facts, where we replace summation by disjunction,
multiplication by conjunction, and probabilities by truth values of
random variables (or facts).  

In  our initial alarm  example \eqref{ex:alarm}, the DNF
corresponding to 
\verb|calls(mary)| contains the worlds shown in
Table~\ref{tab:marycalls}, and thus  is 
\begin{align}\label{eq:dnf-cm}
(b\wedge e&\wedge hm\wedge hj)\vee (b\wedge
e\wedge hm\wedge\neg hj)
\vee (b\wedge\neg e \wedge hm \wedge hj)\\ &\vee (b\wedge\neg e \wedge
hm\wedge\neg hj) \vee (\neg b\wedge e\wedge hm\wedge hj) \vee
(\neg b\wedge e\wedge hm\wedge\neg hj).\nonumber
\end{align}

\paragraph{Forward Reasoning:} Following the definition of the semantics of CP-logic
\citep{vennekens:tplp09}, forward reasoning can be used to build a
tree whose leaves correspond to possible worlds, on which
success probabilities can be calculated. Specifically, the root of the
tree is the empty set, and in each node, one step of forward reasoning
is executed, creating a child for each possible outcome in the case of
probabilistic facts or annotated disjunctions. 
For instance, consider the program
\begin{align*}
& 0.4::\mathtt{draw}.\\
& 0.2::\mathtt{green} ; 0.7::\mathtt{red}
 ; 0.1::\mathtt{blue} \colonminus \mathtt{draw}.
\end{align*}
\begin{figure}
\centering
\includegraphics[scale=0.5]{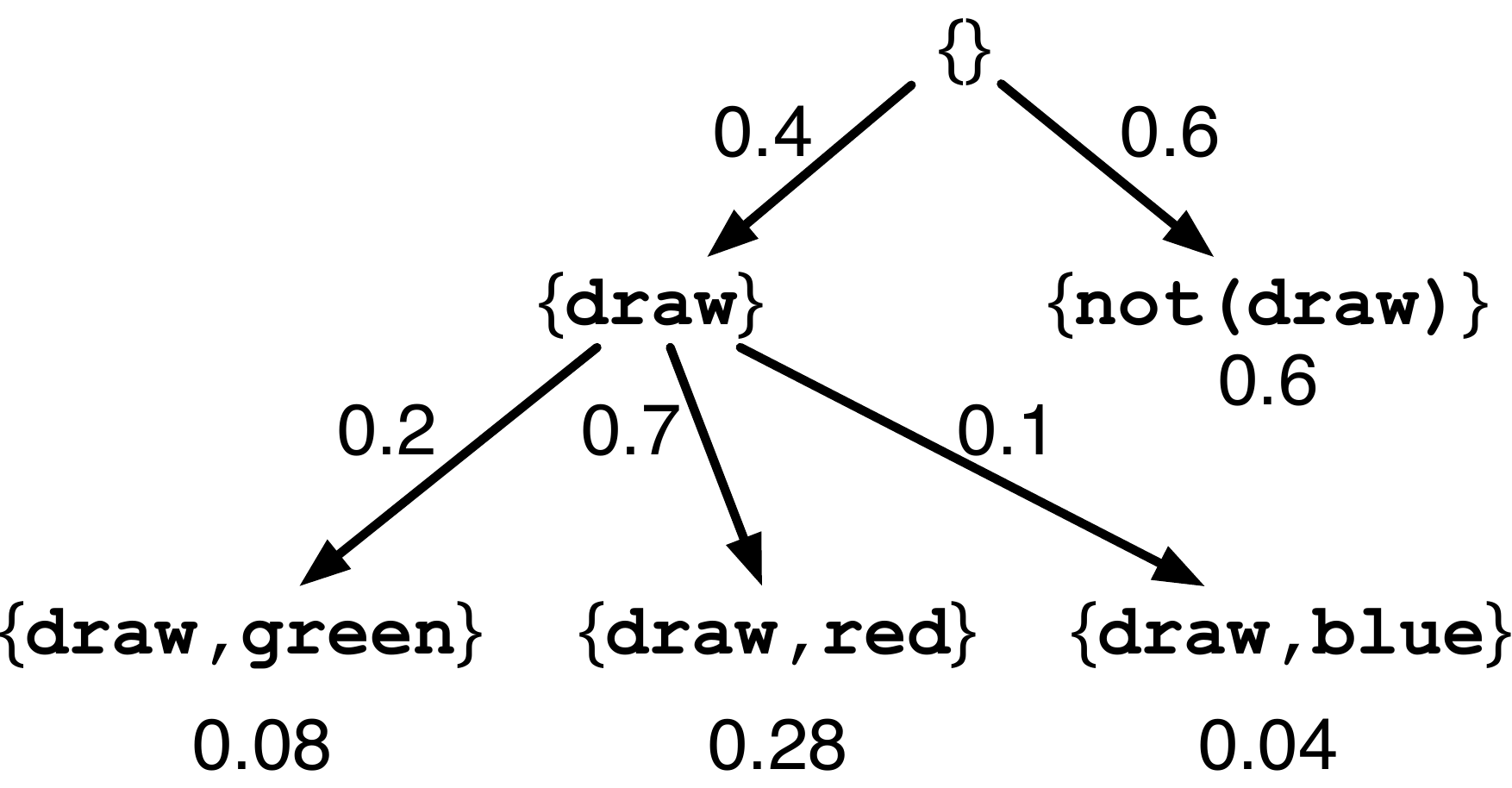}
\caption{Forward reasoning example.}
\label{fig:cptree}
\end{figure}
As illustrated in Figure~\ref{fig:cptree},  the first step
using the probabilistic fact \verb|draw| adds two children to the
root, one containing $\mathtt{draw}$, and
one containing $\mathtt{not(draw)}$.  In the latter case, the body of
the AD is false and thus no further reasoning steps are possible. For
the world where \verb|draw| is true, the AD introduces three children,
adding \verb|green|, \verb|red| and \verb|blue|, respectively, and no
further reasoning steps are possible in the resulting worlds. Thus, each path from the root to a
leaf constructs one possible world, whose probability is the product
of assignments made along the path. Domains for non-ground facts
have to be explicitly provided to ensure termination. 
While this approach clearly
illustrates the semantics, even in the finite case, it suffers from having to
enumerate all possible worlds, and is therefore not used in practice.

\paragraph{Backward Reasoning:} Probably the most common inference strategy in probabilistic logic
programming is to  collect all possible \emph{proofs} or
\emph{explanations} of a given query using backward reasoning,
represent them in a suitable data structure, and compute the
probability on that structure. As discussed in Section~\ref{sec:tasks}, an explanation is a partial
truth value assignment to probabilistic facts that is sufficient to
prove the query via SLD-resolution. For instance, $b\wedge hm$ is the explanation for
\verb|calls(mary)| given by the derivation discussed in Section~\ref{bw-example}
(page \pageref{bw-example}), as it
depends on 
\verb|burglary| and \verb|hears_alarm(mary)| being true, but not on
any particular truth values of \verb|earthquake| and
\verb|hears_alarm(john)|. This query has a second proof, $e\wedge hm$,
obtained by using the first clause for \verb|alarm| during backward
reasoning. We can describe the set of possible worlds where
\verb|calls(mary)| is true by the disjunction of all proofs of the
query, $(b\wedge hm)\vee (e\wedge hm)$, which is more compact than the
disjunction \eqref{eq:dnf-cm} explicitly listing the six possible worlds. We
cannot, however, calculate the probability of this more compact DNF by
simply replacing conjunction by multiplication and disjunction by
addition as we did for the longer DNF above. The reason is that the
two proofs are \emph{not mutually exclusive}, that is, they can be
true in the same possible world. Specifically, in our example this
holds for  the two worlds $b\wedge
e \wedge hm \wedge hj$ and $b\wedge e\wedge hm\wedge\neg
hj$, and the probability
of these worlds, $0.1\cdot 0.2\cdot 0.7\cdot 0.4 + 0.1\cdot 0.2\cdot
0.7\cdot (1-0.4) = 0.014$ is exactly the difference between $0.21$ as
obtained by the direct sum of products $0.1\cdot 0.7+0.2\cdot 0.7$ and the true probability
$0.196$. This is also known as the \emph{disjoint-sum-problem}, which is
\#P-complete~\citep{Valiant79}. Existing languages and systems approach the problem
from different angles. PHA~\citep{Poole92} and
PRISM~\citep{SatoKameya:01} rely on the \emph{exclusive explanation
  assumption}, that is, they assume that the structure of the program guarantees mutual
exclusiveness of all conjunctions in the DNF, which allows one to evaluate it as a
direct sum of products (as done in the PRISM system). This assumption
allows for natural modeling of many models,
including e.g., probabilistic grammars and Bayesian networks, but
prevents direct modeling of e.g., connection problems over uncertain graphs where each edge
independently exists with a certain probability, or simple
variations of Bayesian network models such as our running example. ICL~\citep{Poole00} is closely
related to PHA, but does not assume exclusive explanations. Poole
instead suggests symbolic disjoining techniques to split explanations into
mutually exclusive ones (implemented in AILog2). The ProbLog1
implementation of ProbLog~\citep{DeRaedt07-IJCAIa,Kimmig11} has
been the first probabilistic programming system representing DNFs as Binary Decision
Diagrams (BDDs), an advanced
data structure that disjoins explanations. This technique has
subsequently also been adopted for ICL and
LPADs in the cplint and PITA
systems~\citep{Riguzzi09,Riguzzi11}. AILog2 and cplint also support
computing conditional probabilities. \cite{riguzzi:tcj13} has introduced
an approach called PITA(OPT) that automatically recognizes certain independencies that
allow one to avoid the use of disjoining techniques when computing marginal probabilities. Given its focus on proofs, backward
reasoning can easily be adapted to solve the VIT task of finding most
likely proofs, as done in the PRISM, ProbLog1 and PITA systems.

\paragraph{Reduction to Weighted Model Counting:} A third way to approach the logic side of inference in probabilistic
logic programming has been suggested by \cite{Fierens11,Fierens13}, who use   
the propositional logic semantics of logic
programming  to reduce MARG inference to 
\emph{weighted model counting} (WMC) and MPE inference to weighted MAX-SAT. The first step again builds a
Boolean formula representing all models where the query
is true, but this time, using conjunctive normal form (CNF), and
associating a weight with every literal in the formula. More specifically, it grounds the parts of
the logic program relevant to the query (that is, the rule groundings
contributing to a proof of the query, as determined using backward
reasoning), similar to what happens in
answer set programming, transforms this ground
program into an equivalent CNF based on the semantics of logic
programming, and defines the weight function for the second step using
the given probabilities.  The
second step can then use any existing approach to
WMC or weighted MAX-SAT, such as representing 
 the CNF as an  \emph{sd-DNNF}, a data structure on which WMC can be performed
efficiently. 

For instance, the relevant ground program for \verb|calls(mary)| in
our initial alarm example \eqref{ex:alarm} is 
\begin{align*}
0.1::\mathtt{burglary.} ~~~~&0.7::\mathtt{hears\_alarm(mary).} \\
0.2::\mathtt{earthquake.} ~~~~&\\
\mathtt{alarm}&\colonminus\mathtt{earthquake.}\\
\mathtt{alarm}&\colonminus\mathtt{burglary.}\\
\mathtt{calls(mary)}&\colonminus\mathtt{alarm, hears\_alarm(mary).}\\
\end{align*}
Next, the rules in the ground program are translated to equivalent
formulas in propositional logic, taking into account that their head
atoms can only be true if a corresponding body is true:
\begin{align*}
alarm \leftrightarrow earthquake \vee burglary \\
calls(mary) \leftrightarrow alarm\wedge hears\_alarm(mary)
\end{align*}
The conjunction of these formulas is then transformed into CNF as
usual in propositional logic. The weight function assigns the corresponding probabilities to
literals of probabilistic facts, e.g., $w(burglary)=0.1$, $w(\neg
burglary)=0.9$, and $1.0$ to all other literals, e.g.,
$w(calls(mary))=w(\neg calls(mary))=1.0$. The weight of a model is the
product of all literal weights, and the WMC of a formula the sum of
weights of all its models, which exactly corresponds to the success
probability. Evidence can directly be incorporated by conjoining it
with the CNF. 
Exact MARG inference using this approach is implemented in ProbLog2.

\paragraph{Lifted Inference} is the topic of a lot of research in statistical relational learning today \citep{kersting:ecai12,poole:ijcai03}.  Lifted inference wants to realize probabilistic logic inference
at the lifted, that is, non-grounded level in the same way that resolution realizes this for logical inference.
The problem of lifted inference can be illustrated on the following example (cf.~also \cite{Poole08}):

\begin{align*}
&\mathtt{p}::\mathtt{famous(Y).}\\
&\mathtt{popular(X)} \colonminus \mathtt{friends(X,Y), famous(Y).} 
\end{align*}

In this case $P_s(\mathtt{popular(john)}) = 1 - (1-p)^m$ where $m$ is the number of friends of \verb|john|, that is,
to determine the probability that \verb|john| is popular, it suffices
to know how many friends \verb|john| has. We do not need to
know the identities of these friends, and hence, need not
ground the clauses.   

Various techniques for lifted inference have been 
obtained over the past decade. For instance, \cite{poole:ijcai03}
shows how variable elimination, a standard approach to probabilistic
inference in graphical models,  can be lifted
and \cite{vandenbroeck:ijcai11} studied weighted model counting for first order probabilistic logic using a generalization 
of d-DNNFs for first order logic. Lifted inference techniques are -- to the best of our knowledge -- not yet supported by current
probabilistic logic programming language implementations, which explains why we do not provide more details in this paper. 
It remains a challenge for further work. 
A recent survey on lifted inference is provided by \cite{kersting:ecai12}.

\subsection{Approximate Inference using Bounds}\label{sec:bounds}
As the probability of a set of possible worlds monotonically increases if more
models are added, hard lower and upper bounds on the success probability
can be obtained by considering a subset or a superset of all possible
worlds where a query is true. For instance, let $W$ be the set of
possible worlds where a query $q$ holds. The success probability of
$q$ thus is the sum of the probabilities of all worlds in $W$. If we
restrict this sum to a subset of $W$, we obtain a lower bound, and an
upper bound  if
we sum over a superset of $W$. 
In our example, as \verb|calls(mary)| is true in
$b\wedge e\wedge hm\wedge hj$, but false in  $b\wedge
e\wedge\neg hm\wedge hj$, we have $0.1\cdot 0.2\cdot 0.7\cdot 0.4\leq
P_s(\mathtt{calls(mary)}) \leq 1-(0.1\cdot 0.2\cdot (1-0.7)\cdot
0.4)$. 

In practice, this approach is typically used with the DNF obtained by
backward reasoning, that is, the set of proofs of the query, rather
than with the possible worlds
directly. This has initially been suggested for PHA by \cite{Poole92},
and later also been adapted for ProbLog
\citep{DeRaedt07-IJCAIa,Kimmig08} and LPADs \citep{bragaglia:ilp11}. The idea is to maintain a set of partial
derivations during backward reasoning, which allows one to, at any
point, obtain a lower bound based on all complete explanations or
proofs found so far, and an upper bound based on those together with
all partial ones (based on the assumption that those will become
proofs with probability one). For instance, $(e\wedge hm) \vee b$
provides an upper bound of $0.226$ for the probability of \verb|calls(mary)|
based on the proof $e\wedge hm$ (which provides the corresponding
lower bound $0.14$) and the partial derivation $b$ (which
still requires to prove \verb|hears_alarm(mary)|). Different search
strategies are possible here, including e.g., iterative deepening or
best first search. Lower bounds based on a fixed number of proofs
have been proposed as well, either using 
the $k$ explanations with
highest individual probabilities \citep{Kimmig11}, or the $k$
explanations chosen by a greedy procedure that maximizes the
probability an explanation adds to the one of the current set
\citep{Renkens12}. Approximate inference using bounds is available in
ProbLog1, cplint, and ProbLog2.

\subsection{Approximate Inference by Sampling}\label{sec:sampling}
While probabilistic logic programming  often
focuses on exact inference, approximate inference by sampling is
probably the most popular approach to inference in many other
probabilistic languages. Sampling uses a large number of random executions or randomly
generated possible worlds, from which the probability of a query is 
estimated as the fraction of samples where the query holds. For
instance, samples can be generated by randomly choosing truth
values of probabilistic facts as needed during backward reasoning,
until either a proof is found or all options are exhausted
\citep{Kimmig08,bragaglia:ilp11,riguzzi:fi13}. 
\cite{Fierens13} have used MC-SAT \citep{poon:aaai06} to perform
approximate WMC on the CNF representing all models.  
Systems
for languages that specify generative models, such as
BLOG~\citep{Milch05} and distributional clauses~\citep{Gutmann11},
cf.~Sec.~\ref{sec:dc}, often use forward reasoning to generate
samples. A popular approach to sampling are MCMC algorithms, which,
rather than generating each sample from scratch, generate a sequence
of samples by making random modifications to the previous sample based
on a so-called proposal distribution. This approach has been used
e.g., for the probabilistic functional programming language
Church~\citep{Goodman08}, for BLOG~\citep{arora:uai10}, and for the
probabilistic logic programming languages
PRISM~\citep{sato:ijcai11} and ProbLog~\citep{Moldovan13}. ProbLog1
and cplint provide inference techniques based on backward sampling,
and the PRISM system includes MCMC inference.

\section{Probabilistic Programming Concepts}
\label{sec:concepts}

While probabilistic programming languages based on the distribution
semantics as discussed so far are expressive enough for a wide range
of models, an important part of their power is their support for
additional programming concepts. Based on primitives used in a variety
of probabilistic languages, we discuss a range of such concepts
next, also touching upon their implications for inference.

\subsection{Flexible Probabilities}
A probabilistic fact with \emph{flexible probability} is of the form
$P::atom$ where $atom$ contains the logical variable $P$ that has to be instantiated to a probability when using the fact. The following example models drawing a red ball from an urn with $R$ red and $G$ green balls, where each ball is drawn with uniform probability from the urn:
\begin{equation*}
\begin{array}{ll}
\mathtt{Prob::red(Prob).}\\
\mathtt{draw\_red(R,G)}  \mathtt{\colonminus} & \mathtt{Prob\ is\ R/(R+G),}\\
& \mathtt{ red(Prob).}
\end{array}
\end{equation*}
The combination of flexible probabilities and Prolog code offers a
powerful tool to compute probabilities on-the-fly, cf.~e.g., \citep{Poole08}. 
Flexible probabilities have also been used in extended
SLPs~\citep{Angelopoulos04}, and are supported by the probabilistic
logic programming systems AILog2, ProbLog1, cplint and ProbLog2. 
Indeed, probabilistic facts with flexible probabilities are easily
supported by backward inference as long as
these facts are ground on calling, but cannot directly be used
with exact forward inference, as they abbreviate an infinite set of ground
facts and thus would create an infinite tree of possible worlds.\footnote{If
only finitely many different instances of such a fact are relevant for
any possible world of a given program, a mechanism similarly to the magic set
transformation~\citep{bancilhon:pods86} may circumvent this problem.}

\subsection{Distributional Clauses}
\label{sec:dc}

Annotated disjunctions -- as specified in Section~\ref{sec:ads} -- are
of limited expressivity, as they can only define distributions over a
fixed, finite number of head elements. 
While more flexible discrete distributions can be
expressed using a combination of flexible probabilities and Prolog
code, this may require significant programming
effort. \cite{gutmann:ilp10} introduce Hybrid ProbLog, an
extension of ProbLog to continuous distributions, but their
inference approach based on exact backward reasoning and
discretization severely limits  the use of such distributions. 
 To alleviate these problems, \emph{distributional clauses} were introduced by \cite{Gutmann11}, whom we closely follow.

A \emph{distributional clause} is a clause of the form 
\begin{eqnarray*}
\mathtt{h} \sim {\cal D}  \coloneq \mathtt{b_1,\ \ldots\ ,b_n.} 
\end{eqnarray*}
where $\sim$ is a binary predicate used in infix notation. 
Similarly to annotated disjunctions, the head $ (\mathtt{h}\sim \mathcal{D})\theta$  of a distributional clause
is defined for a grounding substitution $\theta$ whenever $(\mathtt{b_1,\ldots,b_n})\theta$ is true in the semantics of the logic program.  
Then the distributional clause defines the random variable 
$\mathtt{h}\theta$ as being distributed according to the associated
distribution $\mathcal{D}\theta$. Possible distributions include
finite discrete distributions such as a uniform distribution, discrete
distributions over infinitely many values, such as a Poisson
distribution,  and continuous distributions such as Gaussian or Gamma
distributions. The outcome of a random variable $h$ is represented by
the term $\val(h)$. 
Both random variables $h$ and their outcome $\val(h)$ can be used as other terms 
in the program. However, the typical use of  terms $\val(h)$ is inside
comparison predicates such as \texttt{equal/2} or
\texttt{lessthan/2}. 
In this case these predicates act in the same way as probabilistic facts in Sato's distribution semantics. 
Indeed, depending on the value of $\val(h)$ (which is determined
probabilistically) they will be true or false.

Consider the following distributional clause program. 
\begin{align*}
 \mathtt{color(B) \sim discrete((0.7:green),(0.3:blue))} 
         &\coloneq \mathtt{ball(B).}\\  
 \mathtt{diameter(B,MD)\sim gamma(MD1,20) }
         &\mathtt{\coloneq mean\_diameter(\val(color(B)),MD), }\\
& \mathtt{\quad MD1 ~ is ~ MD/20.}\\
    \mathtt{mean\_diameter(green,15).}&\\
    \mathtt{mean\_diameter(blue,25).}&\\
    \mathtt{ball(1). ~ball(2). ~ \ldots ~ ball(k).}&
  \end{align*}
The first clause states that for every ball $\mathtt{B}$, there is a random
variable \verb|color(B)| whose value is either \verb|green| (with
probability $0.7$) or \verb|blue| (with probability $0.3$). This
discrete distribution directly corresponds to the one given by the
annotated disjunction $
\mathtt{0.7::color(B,green); 0.3::color(B,blue)} 
         \coloneq \mathtt{ball(B).}$
The second distributional clause in the example defines a random
variable \verb|diameter(B,MD)| for each ball \verb|B|. This random variable follows a Gamma distribution 
with parameters $\mathtt{MD}/20$ and 20, where the mean diameter
\verb|MD| depends on the color of the ball.

Distributional clauses are 
the logic programming equivalent of the mechanisms employed in
statistical relational languages such as Bayesian Logic (BLOG)
\citep{Milch05}, Church~\citep{Goodman08} and IBAL \citep{Pfeffer01}, which also use programming
constructs to define generative process that can define 
new variables in terms of existing one.

As we have seen in the example, annotated disjunctions can easily be
represented as distributional clauses with finite, discrete
distributions.  
However, distributional clauses are more expressive than annotated
disjunctions (and the standard distribution semantics) as they can also represent continuous distributions.

Performing inference with distributional clauses raises some extra difficulties (see \citep{Gutmann11} for more details). The reason for this 
is that continuous distributions (such as a Gaussian or a Gamma-distribution) 
have uncountable domains. 
Typical inference with constructs such as distributional clauses
will therefore resort to sampling approaches in order to avoid the need for evaluating complex integrals.  
It is quite natural to combine sampling for distributional clauses
with forward reasoning\footnote{Valid distributional clause programs
  are required to have finite support, which ensures termination.}, realizing a kind of generative process, 
though more complex strategies are also possible,
cf.~\citep{Gutmann11}.

\subsection{Unknown Objects}
One of the key contributions of Bayesian Logic (BLOG)~\citep{Milch05}
is that it allows one to drop two common assumptions, namely the
\emph{closed world assumption} (all objects in the world are known in
advance) and the \emph{unique names assumption} (different terms
denote different objects), which makes it possible 
to define probability distributions over outcomes with varying sets of objects. 
This is achieved by defining generative processes that construct possible worlds, where the existence and the properties of objects can depend on objects created earlier in the process.  

As already shown by \cite{Poole08}, such generative processes with an unknown
number of objects can often be modeled using flexible
probabilities and Prolog code to specify a distribution over the
number of objects as done in BLOG. 
Distributional clauses simplify this modeling task, as they make introducing a random
variable corresponding to this number straightforward. We can then use the \texttt{between/3} predicate to enumerate the objects in definitions of predicates that refer to them, cf.~also \citep{Poole08}.  
Below, the random variable \texttt{nballs} stands for the number of
balls, which is Poisson distributed with $\lambda=6$. For each
possible value $\val(\mathtt{nballs})$, the corresponding number of balls are generated which are identified by the numbers $1,2,\ldots,\val(\mathtt{nballs})$. 
\begin{align*} \mathtt{ nballs \sim poisson(6).}\\
\mathtt{ball(N) :- between(1,\val(nballs),N)}.
\end{align*}

\subsection{Stochastic Memoization}
\label{sec:repeat}
A key concept in the probabilistic functional programming language
Church \citep{Goodman08} is \emph{stochastic memoization}. If a random
variable in Church is memoized, subsequent calls to it simply look up
the result of the first call, similarly to \emph{tabling} in logic
programming. On the other hand, for random variables that are not
memoized, each reference to the variable corresponds to an independent
draw of an outcome. In contrast to Church,
probabilistic logic programming languages and their implementations typically do not leave this choice
to the user. 
In ICL, ProbLog, LPADs and the basic distribution semantics as introduced in~\citep{Sato:95}, 
each ground probabilistic fact directly corresponds to a
random variable, i.e., within a possible world, each occurrence of
such a fact has the same truth value, and the fact is thus memoized. Furthermore, the probability of
the fact is taken into account once when calculating the probability
of a proof, independently of the number of times it occurs in that
proof. 
While early versions of PRISM~\citep{Sato:95,sato:ijcai97} used binary or
n-ary probabilistic choices with an argument that explicitly
distinguished between different calls, this argument has been made
implicit later on \citep{SatoKameya:01}, meaning that the PRISM
implementation never memoizes the outcome of a random variable.

The difference between the two approaches can be explained using the following 
example. 
For the AD $( \frac{1}{3}::\mathtt{color(green)} ;
 \frac{1}{3}::\mathtt{color(red)} ;
 \frac{1}{3}::\mathtt{color(blue)})$, there are three answers to the
 goal \texttt{(color(X),color(Y))}, one answer $\mathtt{X}=\mathtt{Y}=\mathtt{c}$ for each color
 $\mathtt{c}$ with probability $\frac{1}{3}$, as exactly one of the facts
 $\mathtt{color(c)}$ is true in each possible world when memoizing color (as in ProbLog and ICL). 
Asking the same question when color is not memoized (as in PRISM) results in $9$ possible answers with
 probability $\frac{1}{9}$ each. The query then -- implicitly --
 corresponds to an ICL or ProbLog query \texttt{(color(X,id1),
   color(Y,id2))}, where the original AD is replaced by a non-ground variant $( \frac{1}{3}::\mathtt{color(green,ID)} ;
 \frac{1}{3}::\mathtt{color(red,ID)} ;
 \frac{1}{3}::\mathtt{color(blue,ID)})$ and 
 \texttt{id1} and \texttt{id2} are trial identifiers that are unique
 to the call.

Avoiding the memoization of probabilistic facts is necessary in order to  
model stochastic automata, probabilistic grammars, or stochastic logic programs \citep{Muggleton:96} under the distribution semantics.
There, a new rule is chosen randomly for each occurrence of the same nonterminal state/symbol/predicate within a derivation, and each such choice contributes to the probability of the derivation. The rules for a nonterminal thus form a family of independent identically distributed random variables, and each choice is automatically associated with one variable from this family.

Consider the following stochastic logic program. It is in fact a fragment of a stochastic definite clause
grammar;  the rules essentially encode the probabilistic context free grammar rules defining $0.3: vp \rightarrow verb$, 
$0.5: vp \rightarrow verb, np$ and $0.2: vp \rightarrow verb, pp$. There are three rules for the non-terminal $vp$
and each of them is chosen with an associated probability. Furthermore, the sum of the probabilities for these rules equals 1.

\begin{align*}
\mathtt{0.3: vp(H,T)}&\colonminus \mathtt{  verb(H,T).}\\
\mathtt{0.5: vp(H,T)}&\colonminus \mathtt{  verb(H,H1),np(H1,T).}\\
\mathtt{0.2: vp(H,T)}&\colonminus \mathtt{  verb(H,H1),pp(H1,T).}\\
\end{align*}

This type of stochastic grammar can easily be simulated in the distribution semantics using 
one {\em dememoized} AD (or switch) for each  non-terminal, a rule
calling the AD to 
make the selection, and a set of rules linking the selection to the
SLP rules:\footnote{The
  $\mathtt{dememoize}$ keyword is used for clarity here; it is not
  supported by existing systems.}
\begin{align*}
&\mathtt{dememoize~~0.3:: vp\_sel(rule1);0.5:: vp\_sel(rule2); 0.2 ::
  vp\_sel(rule3).}\\
&\mathtt{ vp(H,T)}\colonminus \mathtt{  vp\_sel(Rule), vp\_rule(Rule,H,T).}\\
&\mathtt{vp\_rule(rule1,H,T)}\colonminus \mathtt{  verb(H,T).}\\
&\mathtt{vp\_rule(rule2,H,T)}\colonminus \mathtt{  verb(H,H1),np(H1,T).}\\
&\mathtt{vp\_rule(rule3,H,T)}\colonminus \mathtt{  verb(H,H1),pp(H1,T).}\\
\end{align*}

All inference approaches discussed here naturally support stochastic memoization; this
includes the ones implemented in 
AILog2, ProbLog1, ProbLog2, cplint and PITA. The PRISM system uses
exact inference based on backward reasoning in the setting without
stochastic memoization. In principle, stochastic memoization can be
disabled in backward reasoning by automatically adding a unique identifier to each occurrence of
the same random variable. However, for techniques that
build propositional representations different from mutually exclusive DNFs (such as the DNFs of BDD-based methods and the
CNFs when reducing to WMC), care is needed to ensure that these
identifiers are correctly shared among different explanations when
manipulating these formulas. 
Backward sampling can easily deal with both memoized and dememoized
random variables. As only one possible world is considered at any
point, each repeated occurrence of the same dememoized variable is
simply sampled independently, whereas the first result sampled within
the current world is reused for memoized ones. 
Forward sampling cannot be used without stochastic memoization, as it is
unclear up front how many instances are needed. 
MCMC methods have been developed both for ProbLog (with memoization)
and PRISM (without memoization).

\subsection{Constraints}
In knowledge representation, answer set programming and databases, it is common to allow the user to 
specify constraints on the possible models of a theory.  
In knowledge representation, one sometimes distinguishes inductive definitions (such as the 
definite clauses used in logic programming) from constraints.
The former are used to define predicates, the latter impose constraints on possible worlds. 
While the use of constraints is still uncommon in probabilistic logic programming\footnote{Hard   
and soft constraints are used in Markov Logic~\citep{Richardson:06},
but Markov Logic does not support inductive definitions as this requires a least Herbrand semantics, cf. \cite{FierensNIPS12}.
}
it is conceptually easy to accommodate this when working with the distribution semantics, cf. \cite{FierensNIPS12}.
While such constraints can in principle be any first-order logic formula, we will employ clausal constraints here.

A \emph{clausal constraint} is an expression of the form 
\begin{equation*}
\mathtt{h_1 ;\ \ldots\ ; h_N \colonminus b_1,\ \ldots\ ,b_M.}
\end{equation*}
where the $h_i$ and $b_j$ are literals. The constraint specifies that whenever $(b_1 \ldots b_M) \theta$ is true
for a substitution $\theta$ grounding the clause at least one of the $h_i\theta$ must also be true. 
All worlds in which a constraint is violated become impossible, that is, their probability becomes 0.
Constraints are very useful for specifying complex properties that 
possible worlds must satisfy. 

To illustrate constraints, reconsider the alarm example and assume that it models a situation in the 1930s where there 
is only one phone available in the neighborhood implying that at most one person can call. This could be represented by the constraint
\begin{equation*}
\mathtt{X=Y \colonminus calls(X), calls(Y).}
\end{equation*}
Imposing this constraint would exclude all worlds in which both Mary and John hear the alarm and call. 
The total probability mass for such worlds is $0.4 \cdot 0.8 = 0.32$. By excluding these worlds, one looses 
probability mass and thus has to normalize the probabilities of the remaining possible worlds. 
For instance, the possible world corresponding to the truth
value assignment \verb|burglary|=\true, \verb|earthquake|=\false,
\verb|hears_alarm(mary)|=\true, \verb|hears_alarm(john)|=\false\  
yielded a probability mass of 
$0.1\cdot (1-0.2)\cdot 0.7 \cdot (1-0.6) = 0.0336$ without constraints. 
Now, when enforcing the constraint, one obtains $0.0336/(1-0.32)$.
Thus the semantics of constraints correspond to computing conditional probabilities where one conditions
on the constraints being satisfied. 

Handling constraints during inference has not been a focus of
inference in probabilistic logic programming, and  -- to the best of
our knowledge -- no current system provides explicit support for
constraints.

\subsection{Negation as Failure} \label{sec:naf}
So far, we have only considered probabilistic programs using definite
clauses, that is, programs that only use positive literals in clause
bodies, as those are guaranteed to have a unique model for any truth
value assignment to basic probabilistic events. It is however possible
to adopt Prolog's \emph{negation as  failure} on ground literals under
the distribution semantics, as long as all truth values of derived
atoms are still uniquely determined by those of the basic facts, 
cf., e.g.,
\citep{Poole00,sato:ijcai05,KimmigSRL09,Riguzzi09,Fierens13}. Then, in
each possible world, any ground query \verb|q| either succeeds or
fails, and its negation \verb|not(q)|  succeeds in exactly those
worlds where \verb|q| fails. Thus, the
probability of a ground query \verb|not(q)| is the sum of the
probabilities of all possible worlds that do \emph{not}
entail~\verb|q|. 
Consider the following variant of our alarm example, where people also
call if there is no alarm, but they have gossip to share:
\begin{align*}
0.1::\mathtt{burglary.} ~~~~&0.7::\mathtt{hears\_alarm(mary).} \\
0.2::\mathtt{earthquake.} ~~~~&0.4::\mathtt{hears\_alarm(john).}\\
0.3::\mathtt{has\_gossip(mary).} ~~~~&0.6::\mathtt{has\_gossip(john).}\\
\mathtt{alarm}&\colonminus\mathtt{earthquake.}\\
\mathtt{alarm}&\colonminus\mathtt{burglary.}\\
\mathtt{calls(X)}&\colonminus\mathtt{alarm, hears\_alarm(X).}\\
\mathtt{calls(X)}&\colonminus\mathtt{not(alarm), has\_gossip(X).}\\
\mathtt{call}&\colonminus\mathtt{calls(X).}
\end{align*}
The new rule for \verb|calls(X)| can only possibly apply in worlds where
\verb|not(alarm)| succeeds, that is, \verb|alarm| fails, which are
exactly those containing neither \verb|burglary| nor
\verb|earthquake|. Using $gm$ as shorthand for
\verb|has_gossip(mary)|$=\true$, we obtain the additional explanation $\neg
e\wedge \neg b \wedge gm$ for
\verb|calls(mary)|. Thus, in the presence of negation, explanations no
longer correspond to sets of probabilistic facts as in the case of
definite clause programs, but to sets of positive and negative
literals for probabilistic facts. 
While \verb|not(alarm)| has a single explanation
in this simple example, in general, explanations for negative literals
can be much more complex, as they have to falsify every possible
explanation of the corresponding positive literal by flipping the
truth value of at least one probabilistic fact included in the
explanation.

Negation as failure can be handled in forward and backward reasoning
both for exact inference and for sampling, though forward reasoning has to ensure to proceed in
the right order. Exact inference with backward reasoning often
benefits from tabling. Negation as failure complicates approximate
inference using bounds, as explanations for failing goals have to be
considered. AILog2, ProbLog1, ProbLog2, cplint and PITA all support
negation as failure in their exact and sampling based approaches. The PRISM system follows the approach proposed by
\cite{sato:ijcai05} and compiles negation into a definite clause
program with unification constraints. Current MCMC approaches in probabilistic logic
programming do not support negation beyond that of probabilistic facts. 

\subsection{Second Order Predicates}
\label{sec:aggregation}
When modeling relational domains, it is often convenient to reason
over sets of objects that fullfil certain conditions, for instance, to
aggregate certain values over them. In logic programming, this is
supported by second order predicates such as \texttt{findall/3}, which
collects all answer substitutions for a given query in a list.  In the following example, the query \verb|sum(S)| will first collect all arguments of \verb|f/1| into a list and then sum the values using predicate \verb|sum_list/2|, thus returning \verb|S=3|.
\begin{align*}
&\mathtt{f(1).}\\
&\mathtt{f(2).}\\
&\mathtt{sum(Sum)}\colonminus\mathtt{findall(X,f(X),L), sum\_list(L,Sum).}
\end{align*}
Note that in Prolog, the list returned by \verb|findall/3| is unique. Under the distribution semantics, however, this list will be different depending on which possible world is considered. 
To illustrate this, we replace the definition of \verb|f/1| in our example with probabilistic facts:
\begin{align*}
&0.1::\mathtt{f(1).}\\
&0.2::\mathtt{f(2).}\\
&\mathtt{sum(Sum)}\colonminus\mathtt{findall(X,f(X),L), sum\_list(L,Sum).}
\end{align*}
We now have  four sets of facts -- \verb|{f(1),f(2)}|, \verb|{f(1)}|,
\verb|{f(2)}|, and \verb|{}| -- leading to the four possible worlds \verb|{f(1),f(2),sum(3)}|, \verb|{f(1),sum(1)}|,
\verb|{f(2),sum(2)}|, and \verb|{sum(0)}|, as the answer list \verb|L|
is different in each case.

This behavior of second order predicates in the probabilistic setting
can pose a challenge to inference. 
In principle, all inference approaches could 
deal with second order predicates. 
However, exact
approaches would 
suffer from a blow-up, as they have to consider all possible lists of
elements -- and thus all possible worlds -- explicitly, whereas in sampling, each sample only considers
one such list. As far as we know, the only system with some support
for second order predicates is cplint, which allows \verb|bagof| and \verb|setof|
with one of its backward reasoning modules \citep{cplintmanual}.

\subsection{Meta-Calls}
\label{sec:metacalls}
One of the distinct features of programming languages such as Prolog
and Lisp is the possibility to use programs as objects within
programs, which enables meta-level programming. For their
probabilistic extensions, this means reasoning about the probabilities
of queries within a probabilistic program, a concept that is central
to the probabilistic programming language Church, which builds upon a
Lisp dialect~\citep{Goodman08}, and has also been considered with
ProbLog~\citep{Mantadelis11}.  
Possible uses of such a feature include filtering of proofs based on
the probability of subqueries, or the dynamic definition of
probabilities using queries, e.g., to implement simple forms of
combining rules as in the following example, where \verb|max_true(G1,G2)| succeeds with the
success probability of the more likely argument. 
\begin{align*}
&\mathtt{P}::\mathtt{p(P).}\\
&\mathtt{max\_true(G1,G2)} \colonminus
\mathtt{prob(G1,P1),prob(G2,P2),max(P1,P2,P),p(P)}.\\
&\mathtt{\% ~rest~ of~ program ~(omitted)}
\end{align*}
In this section, we will use \verb|prob(Goal,Prob)| to refer to an atom returning the success probability \verb|Prob| of goal \verb|Goal|, that is, implementing Equation~\eqref{eq:p_suc}. Note that such atoms are independent queries, that is, they do not share truth values of probabilistic facts with other atoms occurring in a derivation they are part of. Finally, if the second argument is a free variable upon calling, the success probability of \verb|prob(goal,Prob)| is~1. For the sake of simplicity, we will assume here that the second argument will always be free upon calling.\footnote{This is not a restriction, as \texttt{prob(Goal,c)} is equivalent to \texttt{prob(Goal,P),P=c}.}

We extend the example above with the following program.
\begin{align*}
&0.5::\mathtt{a}.~~~0.7::\mathtt{b}.~~~0.2::\mathtt{c}.\\
&\mathtt{d}\colonminus \mathtt{a,not(b)}.\\
&\mathtt{e}\colonminus \mathtt{b,c}.
\end{align*}
Querying for \verb|max_true(d,e)| using backward reasoning will execute two  calls to \verb|prob/2| in sequence: \verb|prob(d,P1)| and \verb|prob(e,P2)|.  
Note that if multiple calls to  \verb|prob/2|  atoms occur in a proof,
they are independent, i.e., even if they use the same probabilistic facts, those will (implicitly) correspond to different copies of the corresponding random variables local to that specific \verb|prob/2| call. Put differently, \verb|prob/2| \emph{encapsulates} part of our possible worlds. In the example, \verb|b| is thus a different random variable in \verb|prob(d,P1)| and \verb|prob(e,P2)|. 
The reason for this encapsulation is twofold: first, the probability of a goal is not influenced by calculating the probability of another (or even the same) event before, and second, as \verb|prob/2| summarizes a set of possible worlds, the value of a random variable cannot be made visible to the outside world, as it may be different in different internal worlds. Indeed, in our example, \verb|b| needs to be false to prove \verb|d|, but true to prove \verb|e|, so using the same random variable would force the top level query to be unprovable. 
We thus obtain a kind of hierarchically organized world: some probabilistic facts are used in the top level query, others are encapsulated in \verb|prob/2| atoms, whose queries might in turn rely on both directly called probabilistic facts and further calls to \verb|prob/2|.  In our example, \verb|prob(d,P1)| uses random variables corresponding to probabilistic facts \verb|a| and \verb|b|, returning $\mathtt{P1}=0.5\cdot (1-0.7) = 0.15$,  \verb|prob(e,P2)| uses random variables corresponding to probabilistic facts \verb|b| and \verb|c|, returning $\mathtt{P2}=0.7\cdot 0.2 = 0.14$, and  the top level query \verb|max_true(d,e)| uses probabilistic fact \verb|p(0.15)| and has probability $P(\mathtt{more\_likely\_is\_true(d,e)}) = 0.15$.

The probability of a derivation is determined by the probabilities of
the probabilistic facts it uses outside all \verb|prob/2| calls. Those
facts define the possible worlds from the point of view of the top
level query. In those worlds, the random variables of the encapsulated
parts are hidden, as they have been aggregated by
\verb|prob/2|. Returning to our example and abstracting from the
concrete remainder of the program, we observe that for any given pair of goals \verb|g1,g2| and suitable program defining those goals, \verb|max_true(g1,g2)| has exactly one proof: the first two body atoms always succeed and return the probabilities of the goals, the third atom deterministically finds the maximum $m$ of the two probabilities,  and the proof finally uses a single random variable \verb|p(m)| with probability $m$. Thus, the query indeed  succeeds with the probability of the more likely goal. 

Another example for the use of \verb|prob/2| is filtering goals based on their probability:
\begin{align*}
&\mathtt{almost\_always\_false(G)} \colonminus
\mathtt{prob(G,P), P < 0.00001}.\\
&\mathtt{\% ~rest~ of~ program ~(omitted)}
\end{align*}
Note that in contrast to the previous example, this is a purely logical decision, that is, the success probability will be either 0 or 1 depending on the goal $G$.

To summarize, using meta-calls to turn probabilities into usable
objects in probabilistic logic programming is slightly different from the other probabilistic programming concepts considered in this paper: it requires a notion of encapsulation or hierarchical world structure and cannot be interpreted directly on the level of individual possible worlds for the entire program. 

\citet{Mantadelis11}  introduce
MetaProbLog\footnote{\url{http://people.cs.kuleuven.be/~theofrastos.mantadelis/tools/metaproblog.tar.gz},
also supports flexible probabilities, stochastic memoization, and negation
as failure}, a prototype implementation for ProbLog supporting nested meta-calls based on
exact backward inference. As
they discuss, meta-calls
can be supported by any inference 
mechanism that can be suspended to perform inference for the query
inside the meta-call. Such suspending is natural in backward
reasoning, where the proof of a subgoal becomes a call to inference rather
than a continuation of backward reasoning. With forward reasoning,
such non-ground \verb|prob(goal,P)| goals raise the same issues as
other non-ground facts. 
Meta-calls of the form
\verb|prob(goal,P)| 
compute the grounding of \verb|P| as the goal's probability, and using
approximate inference to compute the latter will thus influence the grounding of
such a fact, and therefore potentially also the consequences of this
fact. This may affect the result of inference in unexpected ways, and
it is thus unclear in how far approximation approaches
are suitable for meta-calls.
\cite{Goodman08} state that supporting
meta-calls (or nested queries) in MCMC inference in Church is expected
to be straightforward, but do not provide details.  
Meta-calls are not
supported in AILog2, PRISM, ProbLog1, ProbLog2, cplint and PITA.

\subsection{Time and Dynamics}

Among the most popular probabilistic models
are those that deal with dynamics and time such as Hidden Markov
Models (HMMs) and Dynamic Bayesian Networks.
Dynamic models have received quite some attention within probabilistic logic programming.
They can naturally be represented using logic programs through the addition of an extra
"time" argument to each of the predicates. 
\begin{figure}
\centering
\includegraphics[scale=0.5]{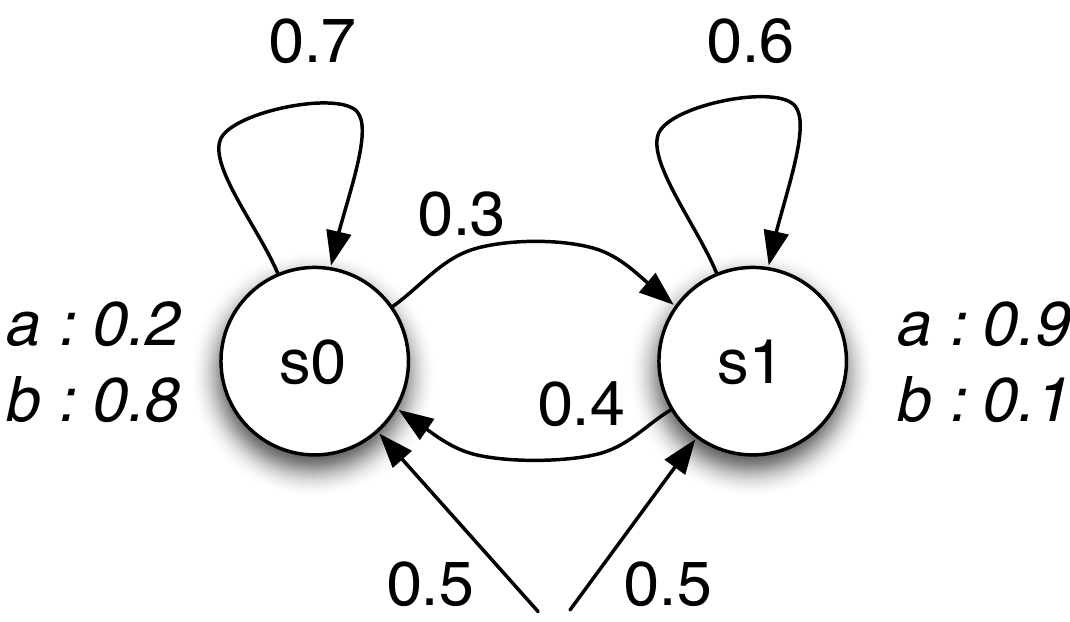}
\caption{Example HMM}
\label{fig:hmm}
\end{figure}
We illustrate this by giving two encodings of the Hidden Markov Model shown in
Figure~\ref{fig:hmm}, where we restrict sequences to a given length
($10$ in the example).  
Following \citet{Vennekens04}, this model can be written as a set of
annotated disjunctions:
\begin{align*}
\mathtt{length(10). }&\\
\mathtt{0.5::state(s0,0); 0.5::state(s1,0).}&\\
\mathtt{0.7::state(s0,T1);
  ~0.3::state(s1,T1)}&{\colonminus}\mathtt{state(s0,T),length(L), T<
  L, T1 ~ is ~ T+1.}\\
\mathtt{0.4::state(s0,T1); ~0.6::state(s1,T1)}&{\colonminus}\mathtt{state(s1,T), length(L), T< L, T1~ is ~T+1.}\\
\mathtt{0.2::out(a,T); ~ 0.8::out(b,T)}&{\colonminus}\mathtt{state(s0,T).}\\
\mathtt{0.9::out(a,T); ~ 0.1::out(b,T)}&{\colonminus}\mathtt{state(s1,T).}\\
\end{align*}

Alternatively, following \citet{sato:ijcai97}, but writing PRISM's
multi-valued switches as unconditional annotated
disjunctions\footnote{In this example, the program structure causes
  the  time argument to 
  act as a unique identifier for different calls to the same AD,
  thus making memoized ADs and dememoized switches equivalent.}, the
model can be written as follows:
\begin{align*}
&0.2::\mathtt{output(s0,a,T)}~\mathtt{;}~0.8::\mathtt{output(s0,b,T).}\\
&0.9::\mathtt{output(s1,a,T)}~\mathtt{;}~0.1::\mathtt{output(s1,b,T).}\\[4pt]
&0.5::\mathtt{init(s0)}~\mathtt{;}~0.5::\mathtt{init(s1).}\\[4pt]
&0.7::\mathtt{trans(s0,s0,T)}~\mathtt{;}~0.3::\mathtt{trans(s0,s1,T).}\\
&0.4::\mathtt{trans(s1,s0,T)}~\mathtt{;}~0.6::\mathtt{trans(s1,s1,T).}\\[4pt]
&\mathtt{length(10).}\\[4pt]
&\mathtt{hmm(List)}\colonminus\mathtt{init(S),~hmm(1,S,List).}\\[4pt]
&\%\mathtt{~last~ time ~T:} \\
&\mathtt{hmm(T,S,[Obs])}\colonminus\mathtt{length(T),~output(S,Obs,T).}\\[4pt]
&\%\mathtt{~earlier~ time ~T:~output~Obs~ in~ state~ S, transit ~from ~S~ to~ Next} \\
&\mathtt{hmm(T,S,[Obs|R])}\colonminus\mathtt{length(L), ~T<L,}\\
&~~~~~~~~~~~~~~~~~~~~~~~~~~\mathtt{output(S,Obs,T),~trans(S,Next,T),}\\
&~~~~~~~~~~~~~~~~~~~~~~~~~~\mathtt{T1 ~is~ T+1,~hmm(T1,Next,R).}
\end{align*}

Forward and backward sampling naturally
deal with a time argument (provided time is bounded in the case of
forward reasoning). Naively using
such a time argument with exact inference results in exponential
running times (in the number of time steps), though
this can often  be avoided using dynamic programming approaches and
principles, as shown by the PRISM system, which achieves the same time
complexity for HMMs as corresponding special-purpose
algorithms~\citep{SatoKameya:01}. 

Other approaches that have devoted special attention to modeling and inference for dynamics
include Logical HMMs \citep{kersting:jair06}, a language for modeling
HMMs with structured states, CPT-L \citep{thon:mlj11}, a dynamic
version of
CP-logic, and the work 
on a particle filter for dynamic distributional clauses \citep{Nitti13}.

\subsection{Generalized Labels}\label{sec:dyna} 
\begin{figure}
\centering
\includegraphics[scale=0.25]{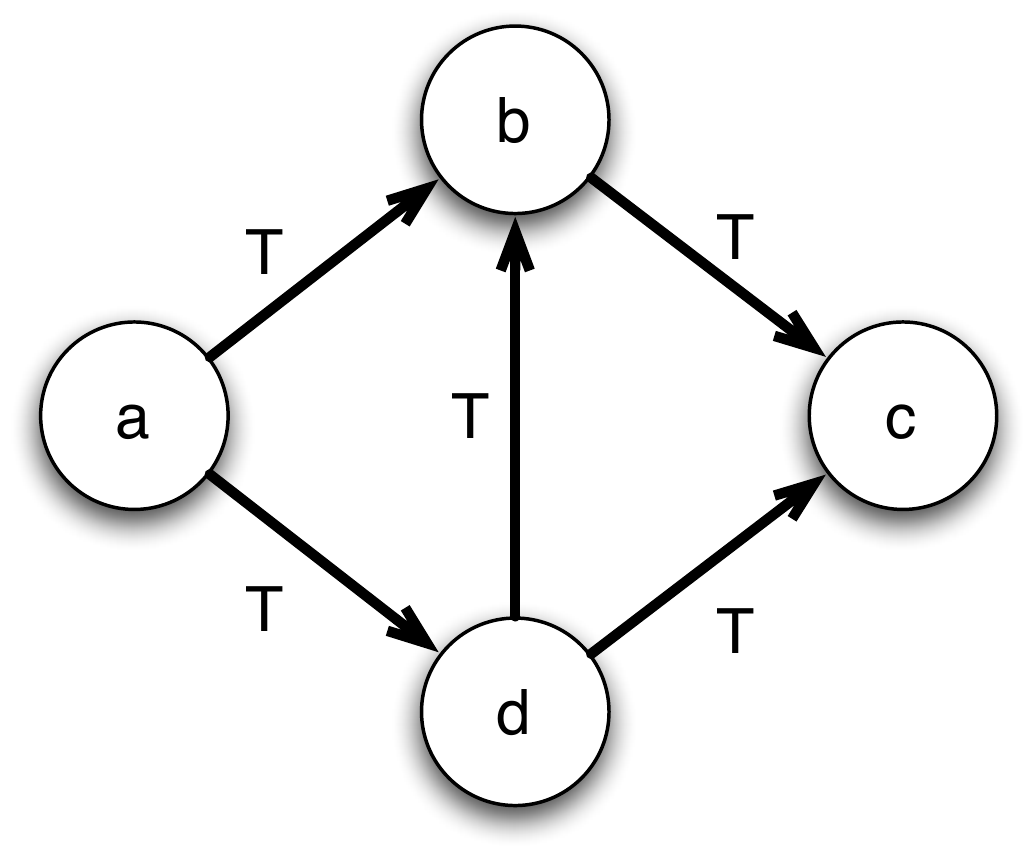}\hspace{2cm}
\includegraphics[scale=0.25]{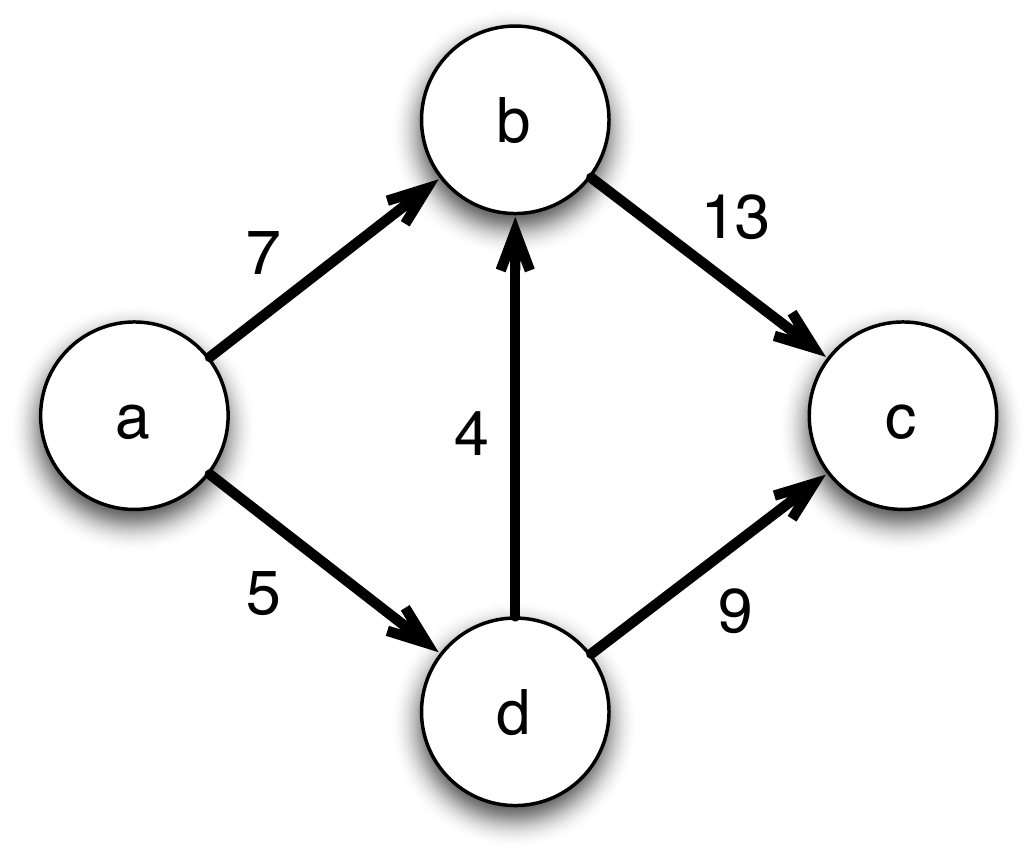}
\caption{Example graph illustrating generalized labels: Boolean case (left),
  shortest path (right).}
\label{fig:dyna}
\end{figure} 
As we have seen in Section~\ref{sec:inference}, computing success probabilities in
probabilistic logic programming is closely related to evaluating the
truth value of a logical formula. Weighted logic programming languages such as
Dyna~\citep{Eisner05}\footnote{Dyna is currently being extended into a
  more general language~\citep{eisner-filardo-2011}, but we consider
  the initial version here, as that one is more closely related to
  the probabilistic programming languages we discuss. } and aProbLog~\citep{kimmig:aaai11} take this
observation a step further and replace probabilities (or Boolean
truth values) by elements from an arbitrary semiring and
corresponding combination operators.\footnote{A \emph{semiring} is a structure $(\mathcal{A}, \oplus, \otimes, 
e^{\oplus}, e^{\otimes})$, where \emph{addition}~$\oplus$ is an associative and commutative binary operation over
the set~$\mathcal{A}$, 
\emph{multiplication}~$\otimes$ is an associative binary operation over
the set~$\mathcal{A}$, 
$\otimes$~distributes
over~$\oplus$, 
$e^{\oplus}\in\mathcal{A}$ is the neutral element
of~$\oplus$, i.e., for all $a\in \mathcal{A}$, $a\oplus e^{\oplus}=a$,
$e^{\otimes}\in\mathcal{A}$ is the neutral element of~$\otimes$, i.e., for all $a\in \mathcal{A}$, $a\otimes e^{\otimes}=a$,
and
for all $a\in \mathcal{A}$, $e^{\oplus}\otimes a = a \otimes
e^{\oplus} = e^{\oplus}$.  In a \emph{commutative semiring}, $\otimes$~is
commutative as well.}

More specifically, Dyna assigns labels to ground facts in a logic
program and computes weights of atoms in the heads of clauses as
follows: conjunction (\verb|,|) in clause bodies is replaced by
semiring multiplication $\otimes$, that is, the weight of a body is
the $\otimes$-product of the weights of its atoms, and if multiple
clauses share the same head atom, this atom's weight is the
$\oplus$-sum of the corresponding bodies, that is, $\colonminus$ is
replaced by semiring addition $\oplus$. 
We illustrate the idea with a logic program defining reachability in a
directed graph adapted from \cite{cohen:iclp08}:
\begin{align*}
\mathtt{reachable(S)}&\colonminus\mathtt{initial(S)}.\\
\mathtt{reachable(S)}&\colonminus\mathtt{reachable(R),edge(R,S)}.
\end{align*}
which in Dyna is interpreted as a system of (recursive) semiring equations
\begin{align*}
reachable(S) ~&\oplus= initial(S).\\
reachable(S) ~&\oplus= reachable(R) \otimes edge(R,S).
\end{align*}
To get the usual logic programming semantics, we can combine this program with facts labeled  with values from
the Boolean semiring (with $\otimes = \wedge$ and $\oplus = \vee$):
\begin{align*}
&\mathtt{initial(a)=T}\\
&\mathtt{edge(a,b)=T ~~~ edge(a,d)=T  ~~~ edge(b,c)=T ~~~ edge(d,b)=T ~~~ edge(d,c)=T}
\end{align*}
which means that the weights of \verb|reachable| atoms are computed as follows:
\begin{align*}
reachable(a) &= initial(a) = T\\
reachable(d) & = reachable(a)\wedge edge(a,d) = T\\
reachable(b) &= reachable(a) \wedge edge(a,b) \vee reachable(d)\wedge
edge(d,b) =  T\\
reachable(c) &= reachable(b)\wedge edge(b,c) \vee reachable(d)\wedge
edge(d,c) = T
\end{align*}
Alternatively, one can label facts with non-negative numbers denoting costs and use $\otimes = +$
and $\oplus = \min$ to describe single-source shortest paths:
\begin{align*}
&\mathtt{initial(a)=0}\\
&\mathtt{edge(a,b)=7 ~~~ edge(a,d)=5  ~~~ edge(b,c)=13 ~~~ edge(d,b)=4 ~~~ edge(d,c)=9}
\end{align*}
resulting in evaluation
\begin{align*}
reachable(a) &= initial(a) = 0\\
reachable(d) & = reachable(a) + edge(a,d) = 5\\
reachable(b) &= \min( reachable(a) + edge(a,b) , reachable(d) + 
edge(d,b)) =  7\\
reachable(c) &= \min( reachable(b)+ edge(b,c) , reachable(d) +
edge(d,c)) = 14
\end{align*}
That is, the values of \verb|reachable| atoms now correspond to the
length of the shortest path rather than the existence of a path. 

Given its origins in natural language processing, Dyna is closely related to PRISM in two aspects. First, it does not memoize labeled
facts, but takes into account their weights each time they appear in a
derivation, generalizing how each use of a rule in a probabilistic
grammar contributes to a derivation. Second, again as in probabilistic
grammars, it sums the weights of all derivations, but in contrast to
PRISM or grammars does not require them to be mutually exclusive to do
so.

 The inference algorithm of basic Dyna as given by
 \cite{Eisner05}\footnote{Implementation available at \url{http://dyna.org/}}
computes weights by forward reasoning, keeping intermediate results in
an agenda and updating them until a fixpoint is reached, though other
execution strategies could be used as well,
cf.~\citep{eisner-filardo-2011}. 

As Dyna, aProbLog~\citep{kimmig:aaai11} replaces probabilistic facts
by semiring-labeled facts, with the key difference that it bases the
labels of derived facts on the labels of their models rather than
those of their derivations, which requires semirings to be commutative. It thus directly generalizes the success
probability \eqref{eq:p_suc_facts} and the possible world DNF
\eqref{eq:model_dnf}. ProbLog inference algorithms based on BDDs or
sd-DNNFs can be directly adapted to aProbLog.\footnote{A prototype implementation of aProbLog
  is included in ProbLog1, cf.~Footnote~\ref{foot:p1}.}

Rather than replacing probabilities with semiring labels, one can
also combine them with utilities or costs, and use the resulting
language for decision making under uncertainty, as done in
DTProbLog~\citep{vandenbroeck:aaai10}.\footnote{An
  implementation of DTProbLog is included in ProbLog1 and ProbLog2,
  cf.~Footnotes~\ref{foot:p1} and \ref{foot:p2}.}

\section{Knowledge-Based Model Construction}
\label{sec:kbmc-bg}
So far, we have focused on probabilistic logic languages with strong
roots in logic, where the key concepts of logic
and probability are unified, that is, a random variable corresponds to
a ground fact (or sometimes a ground term, as in distributional clauses),
and standard logic programs are used to specify knowledge that can be
derived from these facts. In this section, we discuss a second
important group of probabilistic logic languages with 
strong roots in probabilistic graphical models, such as Bayesian or
Markov networks. These formalisms typically use logic as a templating
language for graphical models in relational domains, and thus take a
quite different approach to combine logic and probabilities, also known as
\emph{knowledge-based model construction} (KBMC). Important
representatives of this stream of research include PLPs~\citep{Haddawy:94}, PRMs~\citep{getoor},
BLPs~\citep{Kersting08}, LBNs \citep{fierens:ilp05}, RBNs
\citep{Jaeger97},  \clpbn~\citep{clpbn}, chain logic
\citep{HommersomLucasECML09}, Markov Logic
\citep{Richardson:06}, and PSL \citep{broecheler:uai10}.

In the following, we relate the key concepts underlying the
knowledge-based model construction approach to those discussed in the
rest of this paper. We again focus on languages based on logic
programming, such as PLPs, BLPs, LBNs, chain logic, and \clpbn, but mostly abstract from the specific language. These representation languages are typically designed so that 
implication in logic ("$\colonminus$") corresponds to the direct influence relation 
in Bayesian networks.  The logical knowledge base is then used
to construct a Bayesian network.  So inference proceeds in two steps: 
the logical step, in which one constructs the network, and the 
probabilistic step, in which one performs probabilistic inference on
the resulting network. We first discuss modeling Bayesian networks and their
relational counterpart in the context of the distribution semantics,
and then focus on \clpbn\ as an example of a KBMC approach whose
primitives clearly expose the separation between model
construction via logic programming and probabilistic inference on the
propositional model.

\subsection{Bayesian Networks and Conditional Probability Tables}
\begin{figure}[t]
  \centering
\begin{minipage}[c]{2cm}
    \begin{tabular}{|c|c|}\hline
        C & P(\texttt{s}) \\ \hline
        t & 0.10 \\
        f & 0.50\\ \hline
      \end{tabular}
\end{minipage}
\begin{minipage}[c]{5cm}
\includegraphics[scale=0.3]{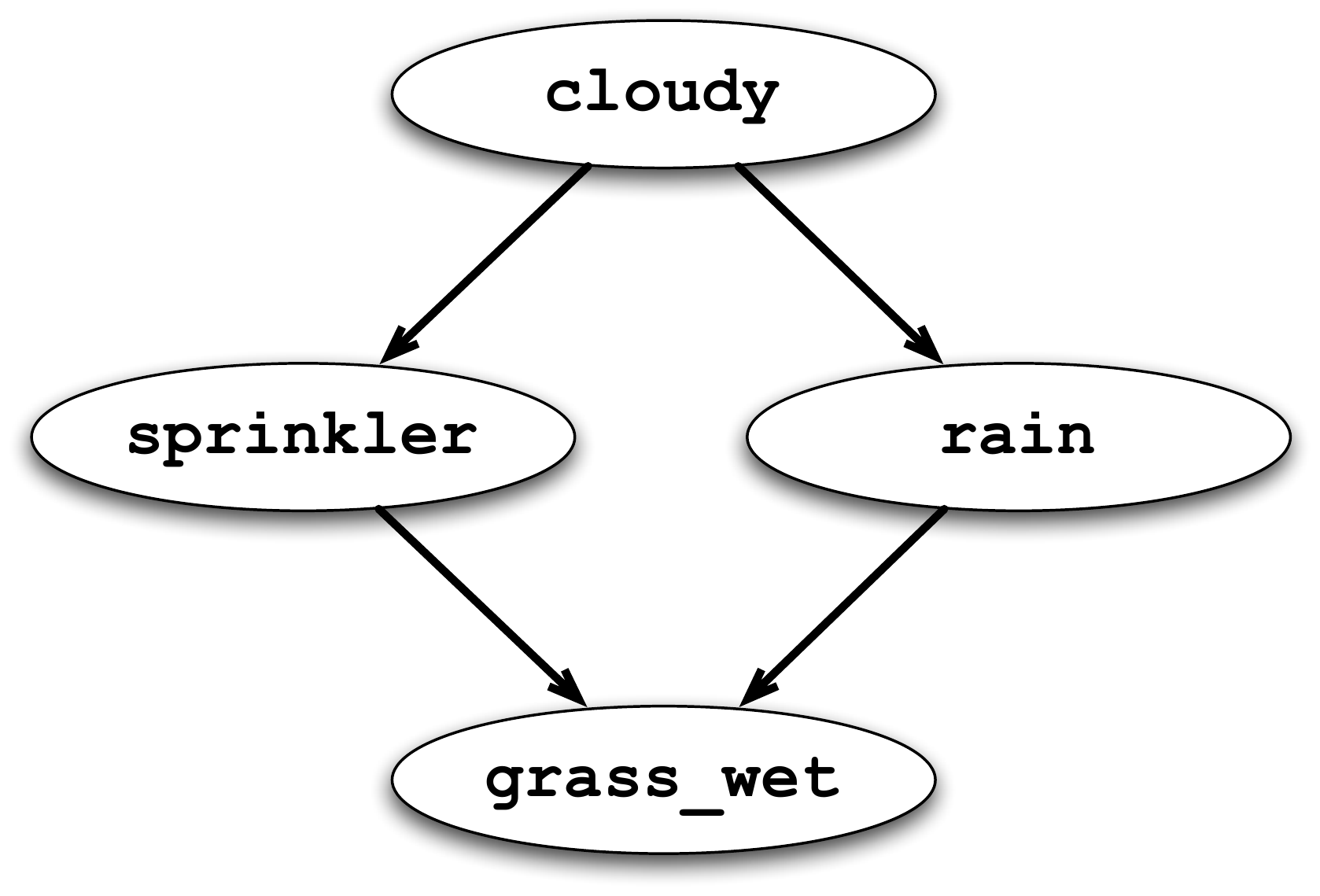}
\end{minipage}
\begin{minipage}[c]{3cm}
\centering
  \begin{tabular}{|c|}
         \hline P(\texttt{c})=$0.50$\\ \hline
       \end{tabular}\\
\vspace{0.2cm}
    \begin{tabular}{|c|c|}\hline
        C & P(\texttt{r}) \\ \hline
        t & 0.80 \\ f & 0.20\\ \hline
      \end{tabular}\\
 \vspace{0.2cm}
    \begin{tabular}{|cc|c|}\hline
        S & R & P(\texttt{g\_w}) \\ \hline
        t  & t &  0.99 \\
        t& f & 0.90 \\
        f & t & 0.90 \\
        f & f & 0.00 \\ \hline
      \end{tabular}
\end{minipage}
 \caption{The sprinkler network is a Bayesian network modeling an environment where both the sprinkler and the rain can cause the grass getting wet~\citep{russel:ai}.}
  \label{fig:sprinkler}
\end{figure}

A Bayesian network (BN) defines a joint probability distribution over a set
of random variables $\mathcal{V} = \{V_1,\ldots,V_m\}$ by factoring it into a product
of conditional probability distributions, one for each variable $V_i$
given its parents $par(V_i)\subseteq\mathcal{V}$. The parent relation
is given by an acyclic directed graph (cf.~Figure~\ref{fig:sprinkler}), where the random variables are
the nodes and an edge $V_i\rightarrow V_j$ indicates that $V_i$ is a
parent of $V_j$. The conditional probability distributions are
typically specified as 
\emph{conditional probability tables} (CPTs), which form the key
probabilistic concept of BNs.  
For
instance, the CPT on the left of Figure~\ref{fig:sprinkler} specifies
that the random variable \verb|sprinkler| takes value $\true$ with
probability $0.1$ (and $\false$ with $0.9$) if its parent \verb|cloudy| is $\true$, and with
probability $0.5$ if \verb|cloudy| is $\false$. 
Formally, a CPT
contains a row for each possible assignment $x_1,\ldots, x_n$ to the
parent variables $X_1,\ldots, X_n$ 
specifying the distribution $P( X |
x_1,\ldots,x_n)$. As has been shown earlier, e.g., by \cite{Poole:93}
and \cite{Vennekens04}, any Bayesian network can be modeled  in
languages based on the distribution semantics by representing  
 every row in a CPT 
as an annotated disjunction
\begin{equation*}
p_1::X(w_1); \cdots; p_k::X(w_k) \colonminus X_1(v_1), \cdots, X_n(v_n)
\end{equation*}
where $X(v)$ is true when $v$ is the value of $X$. The body of this AD
is true if the parent nodes have the values specified in the
corresponding row of the CPT, in which case the AD chooses a value for
the child from the corresponding distribution. As an example,
consider the sprinkler network shown in Figure~\ref{fig:sprinkler}. The CPT for the root node
\texttt{cloudy} corresponds to an AD with empty body
\begin{equation*}
0.5::\mathtt{cloudy(t)} ; 0.5::\mathtt{cloudy(f)}.
\end{equation*}
whereas the CPTs for \texttt{sprinkler} and \texttt{rain} require the
state of their parent node \texttt{cloudy} to be present in the body
of the ADs
\begin{align*}
0.1::\mathtt{sprinkler(t)} ; 0.9::\mathtt{sprinkler(f)}  &\colonminus \mathtt{cloudy(t)}.\\
0.5::\mathtt{sprinkler(t)} ; 0.5::\mathtt{sprinkler(f)}  &\colonminus \mathtt{cloudy(f)}.\\
0.8::\mathtt{rain(t)} ; 0.2::\mathtt{rain(f)} &\colonminus \mathtt{cloudy(t)}.\\
0.2::\mathtt{rain(t)} ; 0.8::\mathtt{rain(f)} &\colonminus \mathtt{cloudy(f)}.\\
\end{align*}
The translation for the CPT of \texttt{grass\_wet} is analogous. 

\subsection{Relational Dependencies}
Statistical relational learning formalisms such as BLPs, PLPs, LBNs and \clpbn\
essentially replace the specific random variables in the CPTs of
Bayesian networks by logically defined random variable templates,
commonly referred to as \emph{parameterized} random variables or
par-RVs for short \citep{poole:ijcai03}, though the actual syntax
amongst these systems differs significantly. We here use annotated
disjunctions to illustrate the key idea. 
For instance, in a propositional setting, the following annotated
disjunctions express that a specific student's grade in a specific
course probabilistically depends on whether he has read the
corresponding textbook or not:
\begin{align*}
0.6::\mathtt{grade(high)} ; 0.4::\mathtt{grade(low)}  &\colonminus \mathtt{reads(true)}.\\
0.1::\mathtt{grade(high)} ; 0.9::\mathtt{grade(low)}  &\colonminus \mathtt{reads(false)}.\\
\end{align*}
Using logical variables, this dependency can directly be expressed for many
students, courses, and books: 
\begin{align*}
0.6::\mathtt{grade(S,C,high)} ; 0.4::\mathtt{grade(S,C,low)}  &\colonminus \mathtt{book(C,B), reads(S,B)}.\\
0.1::\mathtt{grade(S,C,high)} ; 0.9::\mathtt{grade(S,C,low)}  &\colonminus \mathtt{book(C,B), not(reads(S,B))}.\\
\end{align*}

More concretely, the annotated disjunctions express that $P(grade(S,C) = high) = 0.6$ 
if  the student has read the book of the course 
and $P(grade(S,C) = high) = 0.1$ otherwise.  Thus the predicate $\mathtt{grade}$ depends on $\mathtt{book/2}$ and $\mathtt{reads/2}$.  
The dependency holds for all instantiations of the rule, that is, it acts as a template for all persons, courses, and books.  
This is what knowledge-based model construction approaches all share: the logic acts as a template to generate
dependencies (here CPTs) in the graphical model. 
This also introduces a complication that is not encountered in propositional Bayesian networks
or their translation to annotated disjunctions. To illustrate this, let us 
assume the predicate $\mathtt{book/2}$ is deterministic and known. Then the propositional case 
arises when for each course there is exactly one book. The annotated disjunctions then effectively encode
the conditional probability table $P(Grade | Reads)$. However, if there are multiple books, say two,  
for one course, then the above template would specify two CPTs:
one for the first book, $P(Grade | Reads1)$, and one for the second, $P(Grade | Reads2)$. 
In Bayesian networks, these CPTs need to be combined and there are essentially two ways for realizing this. 

The first is to use a so-called \emph{combining rule}, that is, a function that maps these CPTs into a single 
CPT of the form $P(Grade | Reads1, Reads2)$. The most popular combining rule
is noisy-or, for which $P(Grade = high | Reads_1, ...,Reads_n) = 1-\prod_{i=1}^n(1-P(Grade=high |Reads_i=\true))$ where $n$ is the number of books for the course.  
Using annotated disjunctions, this combining rule is obtained
automatically, cf.~Section~\ref{sec:noisy-or-in-lp}. 
In the statistical relational learning literature, this approach is
followed for instance in RBNs and BLPs, and several other combining
rules exist, cf., e.g.,
\citep{Jaeger97,Kersting08,natarajan:icml05}.
While combining rules are an important concept in KBMC, using them in
their general form under the distribution semantics requires one to change the underlying logic,
which is non-trivial. \cite{Hommersom11} introduce an approach that
models these interactions by combining the distribution semantics with
default logic. Alternatively, one could use meta-calls, cf.~Section~\ref{sec:metacalls}.

The second way of dealing with the two distributions uses aggregation. In this way, 
the random variable upon which one conditions grade is the number of books the person read,
rather than the reading of the individual books. This approach is 
taken for instance  in PRMs and \clpbn. In the context of the distribution
semantics, aggregation can be realized within the logic program using second order
predicates, cf.~Section~\ref{sec:aggregation}. For instance, the
following program makes a distinction between reading more than two,
two, one, or none of the books:
\begin{align*}
&0.9::\mathtt{grade(S,C,high)} ; 0.1::\mathtt{grade(S,C,low)} \colonminus \mathtt{nofbooksread(S,C,N), N>2 . }\\
&0.8::\mathtt{grade(S,C,high)} ; 0.2::\mathtt{grade(S,C,low)} \colonminus \mathtt{nofbooksread(S,C,2).}\\
&0.6::\mathtt{grade(S,C,high)} ; 0.4::\mathtt{grade(S,C,low)}  \colonminus \mathtt{nofbooksread(S,C,1).}\\
&0.1::\mathtt{grade(S,C,high)} ; 0.9::\mathtt{grade(S,C,low)} \colonminus \mathtt{nofbooksread(S,C,0).}\\
&\mathtt{nofbooksread}\mathtt{(S,C,N)}\colonminus
\mathtt{findall(B,(book(C,B),reads(S,B)),List),length(List,N).}\\
\end{align*}

\subsection{Example: \clpbn}
\label{sec:cplbbn}
We now discuss \clpbn~\citep{clpbn} in more detail, as it
clearly exposes the separation between model
construction via logic programming and probabilistic inference on the
propositional model in KBMC. \clpbn\ is embedded in
Prolog\footnote{implementation included in YAP Prolog,
  \url{http://www.dcc.fc.up.pt/~vsc/Yap/}} and uses constraint
programming principles to construct a Bayesian network via
logical inference. Syntactically, \clpbn\ extends logic programming
with \emph{constraint atoms} that (a) define random variables together
with their CPTs and (b) establish constraints linking these random
variables to logical variables used in the logic program. The answer
to a query in \clpbn\ is the marginal distribution of the query
variables, conditioned on evidence if available. 
The first
phase of inference in
\clpbn\ uses backward reasoning in the logic program to collect
all relevant constraints in a constraint store, the second phase computes the marginals in the
Bayesian network defined by these constraints. Conditioning on
evidence is straightforward, as it only requires to add the
corresponding constraints to the store.\footnote{The implementation
  adds evidence declared in the input program to the store at compile time.}

Specifically, a \clpbn\ clause (in canonical form) is either a
standard Prolog clause, or has the following
structure:
\begin{align*}
\mathtt{h(A_1,\ldots,A_n,V)} \colonminus \mathtt{body, \{ V = sk(C_1,\ldots,C_t)~~ with ~~CPT \}.}
\end{align*}
Here, \verb|body| is a possibly empty conjunction of logical atoms,
and the part in curly braces is a constraint atom. 
$\mathtt{sk(C_1,\ldots,C_t)}$ is a Skolem term  not occurring in any other clause of
the program (whose arguments $\mathtt{C_i}$ are given via the input
variables $\mathtt{A_j}$ and the logical \verb|body|), and CPT is a
term of the form $\mathtt{p(Values, Table , Parents )}$, where $\mathtt{Values}$ is a list of possible
values for $\mathtt{sk(C_1,\ldots,C_t)}$, $\mathtt{Parents}$ is a
list of logical variables specifying the parent nodes, and $\mathtt{Table}$ the probability table given as  a list of probabilities, where the order
of entries corresponds to the valuations obtained by backtracking over
the parents' values in the order given in the corresponding
definitions. This CPT term can be given either directly or via the use of
logical variables and unification. 

We first illustrate this for the propositional case, using the
following model\footnote{taken from the examples in the \clpbn\
  system} of the sprinkler Bayesian network as given in
Figure~\ref{fig:sprinkler}:\footnote{We include comments 
for better readability, as \clpbn\ swaps rows and columns of CPTs
  compared to the notation in Figure~\ref{fig:sprinkler}.} 
\begin{verbatim}
cloudy(C) :-
        { C = cloudy  with p([f,t],[0.5,0.5],[]) }.

sprinkler(S) :-
        cloudy(C),                % C = f , t  
        { S = sprinkler with p([f,t], [0.5,0.9,    % S = f
                                       0.5,0.1],   % S = t
                               [C])
        }.

rain(R) :-
        cloudy(C),           % C = f , t
        { R = rain with p([f,t], [0.8,0.2,   % R = f
                                  0.2,0.8],  % R = t
                          [C])
        }.

wet_grass(W) :-
        sprinkler(S),
        rain(R),
        { W = wet with p([f,t],
                 /* S/R = f/f, f/t, t/f, t/t */
                         [1.0, 0.1, 0.1, 0.01,  % W = f
                          0.0, 0.9, 0.9, 0.99], % W = t
                         [S,R])
        }.
\end{verbatim}
In the clause for the top node cloudy, the body consists of a single
constraint atom that constrains the
logical variable \verb|C| to the value of the 
random variable \verb|cloudy|. This random variable takes values \verb|f| or \verb|t|
with probability $0.5$ each, and has an empty parent list. Note that
within constraint atoms, the \verb|=| sign does not denote Prolog unification,
but an equality constraint between a logical variable and the value of
a random variable. 
 The clause for
\verb|sprinkler| calls \verb|cloudy(C)|, thus setting up a
constraint between \verb|C| and the \verb|cloudy| random variable, and
then uses \verb|C| as the only parent of the 
random variable \verb|sprinkler| it defines. The first column of the CPT corresponds
to the first parent value, the first row to the first child value, and
so on, i.e., in case of \verb|cloudy=f|, the probability of
\verb|sprinkler=f| is $0.5$, whereas for \verb|cloudy=t|, it is
$0.9$. The other two random variables are defined analogously, with
their clauses again first calling the predicates for the parent
variables to include the corresponding constraints.  To answer the
query \verb|sprinkler(S)|, which asks for the marginal of the random
variable \verb|sprinkler|, \clpbn\ performs backward reasoning to
find all constraints in the proof of the query, and thus the part of
the Bayesian network relevant to compute the marginal. This first
calls \verb|cloudy(C)|, adding the constraint \verb|C=cloudy| to the
store (and thus the \verb|cloudy| node to the BN), and then adds the
constraint \verb|S=sprinkler| to the store, and the \verb|sprinkler|
node with parent \verb|cloudy| to the BN. 

When defining relational models, random variables can be parameterized
by logical variables as in the following clause from the school example
included in the implementation: 
\begin{verbatim}
registration_grade(R, Grade) :-
   registration(R, C, S),
   course_difficulty(C, Dif),
   student_intelligence(S, Int),
   grade_table(Int, Dif, Table),
   { Grade = grade(R) with Table }.

grade_table(I, D, 
     p([a,b,c,d],
/* I,D =  h h   h m  h l  m h  m m  m l  l h  l m l l */
        [ 0.2,  0.7, 0.85, 0.1, 0.2, 0.5, 0.01, 0.05,0.1 , %a
          0.6, 0.25, 0.12, 0.3, 0.6,0.35,0.04, 0.15, 0.4 , %b
          0.15,0.04, 0.02, 0.4,0.15,0.12, 0.5, 0.6, 0.4,   %c
          0.05,0.01, 0.01, 0.2,0.05,0.03, 0.45, 0.2, 0.1 ],%d 
        [I,D])).
\end{verbatim}
Here, \texttt{registration/3} is a purely logical predicate
linking a registration \verb|R| to a course \verb|C| and a
student \verb|S|.  The predicates \texttt{course\_difficulty} and
\texttt{student\_intelligence} define distributions over possible
values \texttt{h(igh)}, \texttt{m(edium)}, and \texttt{l(ow)} for the
difficulty \texttt{Dif} of course \verb|C| and the intelligence 
\texttt{Int} of student \verb|S|, respectively. For each grounding \verb|r| of the variable
\verb|R| in the database of registrations, this clause defines a random
variable \verb|grade(r)| with values \verb|a|, \verb|b|, \verb|c| and \verb|d| that depends on the difficulty of the
corresponding course and the intelligence of the corresponding
student. In this case, the CPT itself is not defined within the
constraint atom, but provided by a Prolog predicate binding it to a
logical variable. 

Defining aggregation using second order predicates is straightforward
in \clpbn, as random variables
and constraints are part of the object level vocabulary. For instance,
the following clause defines the performance level of a student based
on the average of his grades:
\begin{verbatim}
student_level(S,L) :-
  findall(G,(registration(R,_,S),registration_grade(R,G)),Grades),
  avg_grade(Grades,Avg),
  level_table(T),
  { L = level(S) with p([h,m,l],T,[Avg])}.
\end{verbatim}
Here, \verb|avg_grade/2| sets up a new random variable
whose value is the suitably defined average of the grade list \verb|Grades| (and
which thus has a deterministic CPT) and constrains \verb|Avg| to that
random variable, and \verb|level_table| provides the 
list of CPT
entries specifying how the level depends on this average. We refer to
\cite{clpbn} for a discussion of the inference challenges aggregates
raise.

Despite the differences in syntax, probabilistic primitives, and
inference between \clpbn\ and probabilistic extensions of Prolog
following the distribution semantics, there are also many
commonalities between those. As we discussed above, conditional
probability tables can be represented using annotated disjunctions,
and it is thus possible to transform \clpbn\ clauses into Prolog
programs using annotated disjunctions. On the other hand,
\cite{santoscosta:srl09} discuss the relation between PRISM and \clpbn\
based on a number of PRISM programs that they map into \clpbn\ programs.

\section{Probabilistic Programming Concepts and Inference}
\label{sec:ppci}
\begin{table}
\centering
\begin{tabular}{p{1.2cm}||*{8}{p{0.7cm}} p{0.4cm}  }
        & \rotatebox{30}{Flexible Probabilities} 
        &\rotatebox{30}{Continuous Distributions}
        & \rotatebox{30}{Stochastic Memoization} 
         & \rotatebox{30}{Negation as Failure}   
        & \rotatebox{30}{2nd Order Predicates}   
        & \rotatebox{30}{Meta-Calls}   
        & \rotatebox{30}{Time and Dynamics}   
        & \rotatebox{30}{Generalized Labels (aProbLog)}   
\\ \hline \hline
forward exact & no & no  & with & yes& yes$^*$ & no & yes$^*$& yes\\\hline
backward exact & yes & limited & with or without& yes &yes$^*$  & yes&yes$^*$ & yes\\\hline
\hline
forward sampling  & no & yes & with & yes &
yes & no& yes& n.a. \\\hline 
backward sampling& yes & yes & with or without &
yes& yes&yes& yes & n.a.\\\hline \hline
\end{tabular}
\caption{Relation between key probabilistic programming concepts and
  main dimensions of inference; see Section~\ref{sec:ppci} for details. 
  ($^*$ number of proofs/worlds exponential in length of answer list or time sequence)
}
\label{tab:concepts-inference}
\end{table}

We round off this survey by summarizing the relations between the
dimensions of SUCC inference as discussed in Section~\ref{sec:inference} and
the probabilistic programming concepts identified in
Section~\ref{sec:concepts}. On the probabilistic side, we focus on \emph{exact inference} versus
\emph{sampling}, as conclusions for exact inference carry over to
approximate inference with bounds in most cases. On the logical side,
we focus on \emph{forward} 
versus  \emph{backward} reasoning, as conclusions for backward
reasoning carry over to the approach using weighted model counting. 
We provide an overview
in Table~\ref{tab:concepts-inference}, where we omit the concepts
\emph{unknown objects}, as those are typically simulated via flexible probabilities and/or continuous
  distributions, and \emph{constraints}, as those have not yet been considered
  during inference. For \emph{generalized labels}, we focus on
  aProbLog, as it is closer to the distribution semantics than Dyna,
  due to its semantics based on worlds rather than derivations. We do not include MCMC here, as existing MCMC
  approaches in the context of the distribution semantics are limited
  to the basic case of definite clause programs without additional
  concepts. 

\paragraph{Dimensions of inference:} The main difference between exact inference and sampling is that the
former has to consider \emph{all} possible worlds or all proofs of the
query, whereas the latter always considers one possible world or proof
in isolation. As \emph{second order predicates} and \emph{time and dynamics} can
increase the number of proofs exponentially (in the length of the
answer list or the number of time steps), they are more easily handled
by sampling based approaches, though tabling can significantly improve
performance of exact inference in dynamic domains. Sampling based
approaches do not directly apply for \emph{generalized labels}, as sampling
exploits the  probabilistic semantics of fact labels.

The main difference between forward and backward reasoning is that the
former generates all consequences of the probabilistic logic program,
whereas the latter is query-driven and only considers relevant
consequences, which can drastically improve efficiency. This difference is well-known in logic programming, and
becomes even more important in the probabilistic setting, where we are
interested in not just a single world or proof, but in all possible
worlds or all proofs. The fact that backward reasoning is query-driven
makes it well-suited for \emph{flexible probabilities} and
\emph{meta-calls}, which cannot directly be handled in forward reasoning. The
reason is that the corresponding subgoals have an infinite number of
groundings, among which backward reasoning easily picks the relevant
ones, which forward reasoning cannot do. The same effect makes it
necessary to use \emph{stochastic memoization} in forward reasoning, while backward
reasoning can support dememoization (as in PRISM) as well as
memoization (as in the various ICL, ProbLog and LPAD systems). 

The roots of the distribution semantics in logic programming become
apparent when considering inference for the two remaining key
concepts, \emph{negation as failure} and continuous distributions as
provided by \emph{distributional clauses}. While the logic concept of negation as
failure is naturally supported in all combinations of exact inference
or sampling and forward or backward reasoning, the probabilistic
concept of continuous distributions is much more challenging, and only
practical in sampling-based approaches.

\paragraph{Inference approaches:} More specifically, \emph{exact
  inference using forward reasoning} in the form discussed in Section~\ref{sec:exact} can be used for all programs with finitely
many finite worlds, which (a) excludes the use of 
non-ground facts without explicitly given domains, flexible
probabilities, meta-calls and continuous probabilities, and (b) requires 
stochastic memoization. As this approach additionally suffers from
having to enumerate all possible worlds, it is not used in
practice.\footnote{Dyna's exact inference is based on
  forward reasoning, but uses a different type of algorithm that
  propagates value updates using
  forward reasoning based on an agenda of pending updates.}

\emph{Exact inference using backward reasoning} is the most widely supported
inference technique in probabilistic logic programming, provided by
AILog2, PRISM, ProbLog1, cplint, PITA and MetaProbLog. PRISM never
uses stochastic memoization, whereas the other systems always use
it. Only very limited forms of continuous distributions can be
supported, cf.~the work on Hybrid ProbLog \citep{gutmann:ilp10}. All other
concepts can be supported, but implementations differ in the ones they
cover. Negation as failure is supported in all implementations. In
addition, AILog2
and cplint support flexible probabilities, MetaProbLog
supports flexible probabilities and meta-calls, and ProbLog1 supports
flexible probabilities, limited use of continuous distributions
(Hybrid ProbLog) and generalized labels (aProbLog). 
 Approximate inference with bounds using backward reasoning is available in ProbLog1 and
cplint, but restricted to definite clause programs, as the use of
negation as failure complicates proof finding (as discussed in Section~\ref{sec:naf}). As the WMC approach as implemented in ProbLog2 uses backward inference to determine the relevant
grounding, that is, the groundings of clauses that appear in some
proof of a query, the same observations as for exact backward
inference apply in this case as well. ProbLog2 supports flexible
probabilities and negation as failure.

\emph{Forward sampling} in its simplest form as discussed in Section~\ref{sec:exact} can be used with programs whose worlds are all
finite, which excludes the use of non-ground facts without explicitly given domains, flexible
probabilities, and meta-calls, and requires stochastic memoization. In
contrast to exact forward inference, forward sampling does support
continuous distributions, as only one value is considered at a
time. None of the probabilistic logic
programming systems discussed here implement 
forward sampling. 

\emph{Backward sampling} is the most flexible approach and can in principle
deal with all concepts except generalized labels. 
 Backward sampling approaches are
provided by ProbLog1 and cplint, which both support flexible
probabilities and negation as failure. PRISM has a builtin for sampling the
outcome of a query using backward reasoning, but does not use it for
probability estimation.

\section{Conclusions} \label{sec:conclusions}
Probabilistic programming is a rapidly developing field of research 
as witnessed by the many probabilistic programming languages and primitives that have 
been introduced over the past few years.  In this paper, we have attempted to provide a gentle introduction
to this field by focussing on probabilistic {\em logic} programming languages 
and identifying the underlying probabilistic concepts that these languages support.
The same concept (e.g., probabilistic choice) can be realized 
using different syntactic primitives (e.g., switches, annotated
disjunctions, etc.) leading to differences in the probabilistic programming languages. 
Probabilistic programming implementations
not only differ in the primitives they provide
but also in the way they perform probabilistic inference. 
Inference is a central concern in these languages,
as probabilistic inference is computationally expensive. 
We have therefore also presented various probabilistic inference mechanisms and discussed 
their suitability for supporting the probabilistic programming concepts. 
This in turn allowed us to position different languages and implementations,
leading to a broad survey of the state-of-the-art in probabilistic logic programming.

\section*{Acknowledgements}
The authors are indebted to Bernd Gutmann and Ingo Thon
for participating in many discussions, and contributing several ideas  during the early stages of the research that finally led to this paper. 
Angelika Kimmig is supported by the Flemish Research Foundation (FWO-Vlaanderen).

\bibliographystyle{plainnat}
\bibliography{probprog-concepts}

\appendix

\section{Logic Programming Basics}
\label{app:lp}
The basic building blocks of logic programs are \emph{variables} (denoted by strings starting with upper case letters), \emph{constants}, \emph{functors} and
\emph{predicates} (all denoted by strings starting with lower case letters).  A \emph{term} is a variable, a constant, or a functor
\texttt{f} of \emph{arity} $n$ followed by $n$ terms $\mathtt{t_i}$, i.e.,
$\mathtt{f(t_1,...,t_n)}$. 
An \emph{atom} is a predicate $p$ of arity $n$ followed by $n$ terms $\mathtt{t_i}$, i.e.,
$\mathtt{p(t_1,...,t_n)}$. A predicate $p$ of arity $n$ is also written as $p/n$. A \emph{literal} is an
atom or a negated atom $\mathtt{not(p(t_1,...,t_n))}$. A \emph{definite
  clause} is a universally quantified expression of the form $h
\colonminus b_1, ... , b_n$ where $h$ and the $b_i$ are atoms.
$h$ is called the \emph{head} of the clause, and $b_1, ... , b_n$ its
\emph{body}. Informally, the meaning of such a clause is that if all
the $b_i$ are true, $h$ has to be true as well. 
 A
\emph{normal clause}  is a universally quantified expression of the form $h
\colonminus b_1, ... , b_n$ where $h$ is an atom and the $b_i$ are
literals.
If $n=0$, a clause is called \emph{fact} and simply written
as $h$. A \emph{definite clause program} or \emph{logic program} for
short is a finite set of definite clauses. A \emph{normal logic
  program} is a finite set of normal clauses. 
A \emph{substitution} $\theta$ is an expression of the form
$\{V_1/t_1,...,V_m/t_m\}$ where the $V_i$ are different variables and
the $t_i$ are terms. Applying a substitution $\theta$ to an expression
$e$ (term or clause) yields the \emph{instantiated} expression $e\theta$
where all variables $V_i$ in $e$ have been simultaneously replaced by
their corresponding terms $t_i$ in $\theta$. If an expression does not
contain variables it is \emph{ground}. Two expressions $e_1$ and $e_2$ can be \emph{unified} if and only if there are substitutions $\theta_1$ and $\theta_2$ such that $e_1\theta_1 = e_2\theta_2$. In Prolog, unification is written using $=$ as an infix predicate. 

The \emph{Herbrand base} of a logic program is the set of ground atoms that can be constructed using the predicates, functors and constants occurring in the program\footnote{If the program does not contain constants, one arbitrary constant is added.}. Subsets of the Herbrand base are called \emph{Herbrand interpretations}. A Herbrand interpretation is a \emph{model} of a clause $\mathtt{h ~~:- ~~b_1,\ldots ,b_n\ldotp}$ if for every substitution $\theta$ such that all $b_i\theta$ are in the interpretation, $h\theta$ is in the interpretation as well. It is a model of a logic program if it is a model of all clauses in the program. The model-theoretic semantics of a definite clause program is given by its smallest Herbrand model with respect to set inclusion, the so-called \emph{least Herbrand model} (which is unique). We say that a logic program $P$ \emph{entails} an atom $a$, denoted $P\models a$, if and only if $a$ is true in the least Herbrand model of $P$.   

The main inference task of a logic programming system is to determine
whether a given atom, also called \emph{query} (or \emph{goal}), is true in the least
Herbrand model of a logic program. If the answer is yes (or no), we also say that the query \emph{succeeds} (or \emph{fails}). If such a query is not ground, inference asks for the existence of an \emph{answer substitution}, that is, a substitution that grounds the query into an atom that is part of the least Herbrand model.

Normal logic programs use the notion of \emph{negation as failure}, that is, for a ground atom $a$, $not(a)$ is true exactly if $a$ cannot be proven in the program. They are not guaranteed to have a unique minimal Herbrand model. Various ways to define the canonical model of such programs have been studied; see, e.g., \cite[Chapter 3]{lloyd:book89} for an overview.

\section{Annotated Disjunctions and Probabilistic Facts}
\label{app:adpf} 
As mentioned in Section~\ref{sec:ads}, each annotated disjunction can be equivalently
represented using a set of probabilistic facts and deterministic
clauses.  Using probabilistic facts is not sufficient, as those
correspond to independent random variables. For instance, using
probabilistic facts 
\begin{equation*}
\begin{array}{lllll}
\frac{1}{3} :: \mathtt{color(green)\ldotp} & ~~~~ &
\frac{1}{3} :: \mathtt{color(red)\ldotp} & ~~~~ & 
\frac{1}{3} :: \mathtt{color(blue)\ldotp} 
\end{array}
\end{equation*}
the probability of \verb|color(green)|, \verb|color(red)| and
\verb|color(blue)| all being true is $1/27$, whereas it is $0$ for the
annotated disjunction $\frac{1}{3}::\mathtt{color(green)} ;
\frac{1}{3}::\mathtt{color(red)} ;
\frac{1}{3}::\mathtt{color(blue)}$. 
On the other hand, we can exploit the fact that negation of
probabilistic facts is easily handled under the distribution
semantics\footnote{For a probabilistic fact \texttt{p::f}, \texttt{not(f)}
  succeeds in a possible world exactly if \texttt{f} is not among the
  probabilistic facts included in that world;
  cf.~Section~\ref{sec:naf} for a more general discussion of negation.}
to encode an AD by simulating a
sequential choice mechanism\footnote{used, e.g., by
  \cite{sato:ijcai97} with parameters learned from data}. With this encoding, the three possible outcomes are mutually
exclusive as in the AD and exactly one will be true in any possible
world: 
\begin{alignat*}{2}
&\begin{array}{lllll}
\frac{1}{3} :: \mathtt{sw\_1(color(green))\ldotp} & ~~~~ &
\frac{1}{2} :: \mathtt{sw\_1(color(red))\ldotp} & ~~~~ & 
1 :: \mathtt{sw\_1(color(blue))\ldotp} 
\end{array}\\
&\begin{array}{lll}
\mathtt{color(green)}  & \colonminus & \mathtt{sw\_1(color(green))}\ldotp\\
\mathtt{color(red)}     & \colonminus & \mathtt{not(sw\_1(color(green))),  sw\_1(color(red))}\ldotp\\
\mathtt{color(blue)}    & \colonminus & \mathtt{ not(sw\_1(color(green))),  not(sw\_1(color(red))), sw\_1(color(blue))}\ldotp\\
\end{array}
\end{alignat*}
Note that the probabilities have been adapted to reproduce the
probabilities of the different head atoms; we discuss the details of
this adaptation below.\footnote{This
  transformation is correct for computing success probabilities, but
  care has to be taken to accomodate for the additional random variables in
  MPE inference.}
This mapping follows the general idea of representing a probabilistic
model in an \emph{augmented space} where random variables can be
assumed independent, while capturing the dependencies in the
deterministic part of the program~\citep{Poole10}.

For non-ground ADs, all logical variables have to be included in the
probabilistic facts to ensure that all groundings correspond to
independent random events. For instance, the AD
$( \frac{1}{2}::\mathtt{color(green)} ;
\frac{1}{2}::\mathtt{color(red)}) \colonminus \mathtt{ball(Ball)}$
would be represented as 
\begin{alignat*}{2}
&\begin{array}{lll}
\frac{1}{2} :: \mathtt{sw\_1(color(green),Ball)\ldotp} & ~~~~ &
1 :: \mathtt{sw\_1(color(red),Ball)\ldotp} 
\end{array}\\
&\begin{array}{lll}
\mathtt{color(green)}  & \mathtt{\colonminus} &\mathtt{ball(Ball),sw\_1(color(green),Ball)}\ldotp\\
\mathtt{color(red)}     & \mathtt{\colonminus} & \mathtt{ball(Ball),not(sw\_1(color(green),Ball)),  sw\_1(color(red),Ball)}\ldotp
\end{array}
\end{alignat*}

As this example suggests,
annotated disjunctions can be expressed using probabilistic facts by representing each annotated disjunction using the set of probabilistic facts $\tilde p_i::sw\_id(h_i,v_1,\ldots,v_f)$ and  the following clauses 
\begin{align}
\label{eq:cswitch}
h_i :\!\!-\ b_1, \cdots , b_m, &not(sw\_id(h_1,v_1,\ldots,v_f)), ..., not(sw\_id(h_{i-1},v_1,\ldots,v_f)),\nonumber\\ &sw\_id(h_i,v_1,\ldots,v_f)
\end{align}
where $id$ is a unique identifier  for a particular AD and $v_1,\ldots,v_f$ are the free variables in the body of the AD. The probability $\tilde p_1$ is defined as $p_1$ and for $i>1$ it is
\begin{equation}
\label{eq:cswitchp}
\tilde p_i \colonequals \begin{cases}
  p_i \cdot\left (1- \sum_{j=1}^{i-1} p_j\right) ^{-1} &\text{if } p_i>0\\
  0  & \text{if }p_i=0
  \end{cases} \enspace .
\end{equation}
One can recover the original probabilities from $\tilde p$ by setting $p_1 \colonequals \tilde p_1$ and iteratively applying the following transformation for $i=2,3,\ldots,n$
\begin{equation}
\label{eq:cswitchpback}
p_i \colonequals  \tilde p_i \cdot\left (1- \sum_{j=1}^{i-1} p_j\right)\enspace.
\end{equation}
Equation~\eqref{eq:cswitchp} and \eqref{eq:cswitchpback} together define a bijection between $p$ and $\tilde p$ which allows one to use parameter learning in either representation and map learned probabilities onto the other representation. If the $p_i$ sum to 1, it is possible to drop the last probabilistic fact $sw\_id(h_n)$ since its probability $\tilde p_n$ is 1.

\end{document}